\documentclass[unnumsec,webpdf,modern,medium]{template}

\onecolumn
\usepackage{fancyhdr,amsmath,amssymb, amsthm,mathtools}
\usepackage{bm}
\usepackage{dsfont}
\usepackage{graphicx}
\usepackage{algorithm}
\usepackage[export]{adjustbox}
\usepackage[british]{babel}
\usepackage{calc} 
\setcitestyle{round}
\usepackage{adjustbox}
\usepackage{hyphenat}
\usepackage{etoolbox}
\usepackage{subcaption}
\usepackage{tabularx}

\makeatletter
\patchcmd{\@@address}
  {\hbox}
  {\relax}
  {}{}
\renewcommand{\addressandsep}{\par}
\makeatother

\newlength{\leftcolumnwidth}
\newcounter{globalrem}

\newcounter{globaldef}

\newcounter{globalprop}

\newcounter{globalem}

\newcounter{globatheo}

\newcounter{globalcor}

\newcounter{globalas}

\begin{document}
\copyrightyear{}
\pubyear{}
\appnotes{$_.$}

\firstpage{1}
\setcounter{secnumdepth}{3}

\fontsize{9.92}{10.48}\selectfont

\title[Low-rank bilinear autoregressive models for three-way criminal activity tensors]{Low-rank bilinear autoregressive models for three-way criminal activity tensors}

\author[1]{Gregor Zens}
\author[2]{Carlos D\'iaz}
\author[3]{Daniele Durante}
\author[4]{Eleonora Patacchini}

\authormark{Zens, D\'iaz, Durante and Patacchini}

\address[1]{\orgdiv{International Institute for Applied Systems Analysis}, \orgname{IIASA},  \country{Laxenburg, Austria}}
\address[2]{\orgdiv{Millennium Nucleus on Criminal Complexity and School of Economics and Business},  \orgname{Universidad Alberto Hurtado},  \country{Santiago, Chile}}
\address[3]{\orgdiv{Department of Decision Sciences and Bocconi Institute for Data Science and Analytics}, \orgname{Bocconi University},  \country{Milan, Italy}}
\address[4]{\orgdiv{Department of Economics},  \orgname{Cornell University},  \country{ Ithaca, NY, US}}

\corresp[]{Gregor Zens, International Institute for Applied Systems Analysis (IIASA), Schlossplatz 1, 2361 Laxenburg, Austria. Email: \href{email: zens@iiasa.ac.at}{zens@iiasa.ac.at}}

\abstract{\fontsize{9.6}{10}\selectfont
Criminal activity data are typically available via a three-way tensor encoding the reported frequencies of different crime categories across time and space. The challenges that arise in the design of interpretable, yet realistic, model-based representations of the complex dependencies within and across these three dimensions have led to an increasing adoption of black-box predictive strategies. While this perspective has proved successful in producing accurate forecasts guiding targeted interventions, the lack of interpretable model-based characterizations of the dependence structures underlying criminal activity tensors prevents from inferring the cascading effects of these interventions across the different dimensions. We address this gap through the design of a low-rank bilinear autoregressive model which achieves comparable predictive performance to black-box strategies, while allowing interpretable inference on the dependence structures of reported criminal activities across crime categories, time and space. This representation incorporates the time dimension via an autoregressive construction that accounts for spatial effects and dependencies among crime categories through a separable low-rank bilinear formulation. When applied to Chicago police reports, the proposed model showcases remarkable predictive performance and also reveals interpretable dependence structures unveiling fundamental crime dynamics. These results facilitate the design of more refined intervention policies informed by the cascading effects of the policy itself.
}

\keywords{Bayesian statistics, criminal activity data, graphical models, matrix-variate time series}

\maketitle
\fontsize{9.92}{10.47}\selectfont
\vspace{-10pt}

\section{\large 1. Introduction} \label{sec_1}
The empirical analysis of criminal activities plays an important role in both theory development and public policy considerations. Theoretical models of crime indicate that criminal incidents do not occur randomly, but rather exhibit complex interdependencies across spatial areas, time, and crime categories \citep[e.g.][]{brantingham2016geometry, weisburd2015law}. As a result, several methods have been proposed to characterize the intricate system of interdependencies within and across these three dimensions. Available solutions  range from interpretable model-based representations \citep[see, e.g.][]{balocchi2023crime, balocchi2019spatial, clark2018modeling,law2014bayesian, liu2017joint, quick2018crime,shoesmith2013space,vicente2023multivariate} to more sophisticated black-box predictive strategies \citep[see, e.g.][]{berk2021artificial, chun2019crime,guo2012understanding,liang2022crimetensor, mandalapu2023crime, walczak2021predicting,  wang2025mragnn,zhang2022interpretable}. Although statistical models allow, in principle, for interpretable inference on the complex dependence structures underlying the observed criminal activity tensors, the validity of the final conclusions is typically undermined by the challenges that arise in the design of sufficiently flexible, yet interpretable,  model-based representations for these dependence structures. As such, the existing statistical models often constrain the analysis by focusing on two dimensions of the three-way tensor, fixing the third  at a selected crime category, a single spatial location, or a pre-specified time interval. Although this perspective has led to key advancements, constraining the analysis only to two dimensions fails to characterize the full set of interdependencies underlying the criminal activity tensor, thereby limiting the policy impact of the resulting inferences. Addressing these issues would require extending the available models for two dimensions to similarly effective formulations for the entire   three-way tensor. In fact, instead of exploring such a direction, most efforts in the recent years have been devoted to the development and implementation of black-box strategies within a predominantly prediction-oriented framework \citep[][]{berk2021artificial,mandalapu2023crime}. These strategies have proved successful in producing accurate forecasts guiding intervention policies, such as {\em hot spot} policing \citep{braga2014effects,weisburd2014hot}, but generally lack a model-based construction, and therefore,  make it challenging to derive meaningful structural insights on the cascading spillover effects of specific localized interventions \citep[e.g.][]{blattman2021place,weisburd2006does}. Learning these effects is fundamental to devise improved intervention policies that not only rely on the forecasted criminal activities, but are also guided by the expected cascading effects of the planned policy itself.

Motivated by the above discussion, we design in Section~\ref{sec:model} a model-based Bayesian representation of three-way criminal activity tensors that achieves a predictive performance comparable to routinely used black-box strategies, while allowing for interpretable inference on the complex dependence structures within and across crime categories, spatial locations and time intervals. This formulation is guided by qualitative and quantitative theories of crime discussed in Section~\ref{sec:lit}, and relies on a temporal autoregressive construction, which incorporates spatial effects and graphical dependencies among different crime categories via a separable low-rank bilinear formulation inspired by the general class of multilinear tensor regression models in \citet{hoff2015multilinear}. Besides reducing the dimensionality of the problem, this novel perspective to the analysis of crime activity tensors crucially unveils inter-temporal dependencies between crime categories shared across spatial locations as well as  among these spatial locations common over crime categories. Such a separable structure is also considered for the noise component of the model, thereby further uncovering across-categories and across-locations immediate reactions to shocks, possibly arising from targeted {\em hot spot} interventions. As clarified within Section~\ref{sec_3}, inference under the proposed statistical model is performed through a Bayesian approach leveraging a suitably designed Gibbs sampling algorithm. The advantages of our novel formulation are illustrated in Section~\ref{sec:application} via an application to a three-way criminal activity tensor obtained from the Chicago Police Department. Besides achieving a predictive performance comparable to routinely implemented black-box strategies, the proposed model unveils yet unexplored statistical interdependencies between the crime categories and among spatial locations over time that augment the set of information available to those policymakers seeking to understand and prevent complex patterns in criminal activities. As discussed in the concluding remarks in Section~\ref{sec:concl}, our novel approach can be applied in general to criminal activity tensors from any urban environment for which similar data are available.

\section{\large 2. Qualitative and quantitative evidence on criminal activities} \label{sec:lit}

\subsection{\large 2.1. Determinants and mechanisms}
\label{det}
Crime is mainly an urban phenomenon. Urban agglomerations typically exhibit higher crime rates compared to smaller towns and rural areas \citep[e.g.][]{glaeser1999}. Although larger cities are often associated with higher economic development, and hence more attractive, this is not the only dimension that explains the distribution of crime across cities. In fact, crime reports are typically sorted unequally over urban locations, with offenses often concentrated by type in relatively small areas. This spatial heterogeneity, coupled with specific temporal persistences, has led social scientists to focus on local determinants rather than individual drivers. 

Three traditional criminology perspectives are useful for understanding the connections among time, space and crime categories. Crime opportunity theories, which focus on the factors shaping space-time crime patterns, particularly the clustering of criminal activities \citep[e.g.][]{brantingham2017notes, cohen1979social}; social control theories, where the structural dimensions of a neighborhood influences residents propensity to exert informal social control \citep{sampson1989community, shaw1942juvenile}; and theories that simply  highlight how local characteristics (e.g. medical marijuana dispensaries, parks, neighborhood stability and socioeconomic disparities) affect incentives for committing a specific crime within a given area \citep{contreras2017block, contreras2020drugs, kubrin2016fringe}. It is important to note that crime clusters are not driven by a single factor in isolation \citep{hipp2020advances}. Rather, the interplay between opportunity, social controls, and local conditions causes urban areas to experience varying levels of temporal exposure to specific types of crimes and violence.

To illustrate the complexity of the determinants and mechanisms linking time, space, and crime categories it is possible to recall the routine activity theory. Originally developed by \cite{cohen1979social}, this perspective offers one of the clearest frameworks for explaining the relationship between opportunity and rational choice \citep{becker1968crime, ehrlich1973participation}. According to this theory, crime is unevenly distributed \citep{weisburd2015law} and occurs when three conditions converge in time and space, namely motivated offenders, suitable targets (valuable objects or potential victims), and the absence of effective interventions to prevent crime \citep{felson2018crime}. As a result, the location and persistence of crime in urban areas are not random but rather shaped by specific situational contexts, which influence how offenders perceive the costs and benefits of criminal activity \citep[][]{becker1968crime}. Crime has higher chance to occur when offenders access potential targets without interference, and since this convergence tends to vary by crime category, distinct spatio-temporal patterns emerge for different offenses. Therefore, the routine activity approach offers a valuable framework to motivate and support the design of models capable of disentangling the complex system of interdependencies in the three-way criminal activity tensor.

Although it is challenging to separate time and space when analyzing urban crime patterns, the temporal dimension of crime determinants has been documented to play a crucial role in shaping its spatial distribution, alongside factors related to the offender behavior. Several criminogenic features of a spatial area tend to remain stable over time, contributing to the persistence of specific criminal activities \citep{johnson2008repeat}. For example, neighborhood characteristics that either promote or inhibit social control are likely to remain constant \citep[e.g.][]{sampson1989community}. In addition, the success of previous offenses often motivates criminals to continue operating in the same area, capitalizing on the knowledge acquired about the available targets and the absence of countermeasures \citep[e.g.][]{cohen1979social}. The literature on near repeat victimization further suggests that offenders are more likely to target locations where similar crimes have been successfully committed in the past \citep{johnson2004burglary, townsley2003infectious}. However, not all crimes are equally susceptible to temporal persistence and spatial clustering \citep[][]{amemiya2020near,chainey2021comparison, chainey2022homicide,haberman2012predictive,johnson2004burglary,johnson2009offender,pease1998repeat,ratcliffe2008near}. In fact, the clustering of urban crime over time and space is shaped by patterns that are heterogeneous across crime categories. In specific urban areas, certain types of offenses are more prevalent than others, leading to a localized concentration of particular crimes (e.g.\ property crimes versus violent crimes). These patterns are shaped by the spatial distribution of suitable targets and by the presence of deterrence interventions \citep{cohen1979social}, which can favor the occurrence of certain crimes over the others. For example, the \textsc{covid}-19 pandemic highlighted the role of crime opportunity in explaining the global reduction in street crime \citep{nivette2021global, trajtenberg2024heterogeneous}, while, at the same time, the literature documented increments in domestic violence and cybercrime \citep{piquero2021domestic, ravindran2023unintended}. Empirical evidence suggests that offenders may have increased the use of violence to compensate for the reduced opportunities brought on by the pandemic, thus reshaping the spatial distribution and clustering of  violent and non-violent crimes \citep{diaz2022stay}. Another example of crime-type clustering heterogeneity results from the strategic decision-making of organized crime groups when selecting or relocating to new territories \citep{varese2011mafia}. These groups typically target urban communities that offer favorable conditions for the development of illegal markets, display socio-economic deprivation, and demonstrate a higher degree of homophily \citep[][]{campana2024organised, katz2011neighborhood}. For instance, gang members are more likely to settle in communities sharing similar social and demographic features. This dynamic helps in explaining why activities such as production, trafficking, and governance \citep[][]{breuer2023structure, campana2018organized}, along with related criminal behaviors, are more prevalent in certain urban areas than in others.

\subsection{\large 2.2. Policy implications}
\label{pol}
The primary policy implication of the geographical concentration and the temporal persistence of crime mainly involves the adoption of targeted deterrence strategies, with emphasis on policing. As most crimes are concentrated in small  areas \citep[][]{telep2018, weisburd2004trajectories}, law enforcement approaches are often tailored to this reality. One of the most popular strategies is the {\em hot spots} approach \citep[][]{braga2014effects,weisburd2014hot}, where police resource allocation focuses on geographic areas with a higher number of crimes than the city's average or where  people are at a higher risk of victimization. In this way, police resources could be allocated to just 5\% of its block segments for addressing more than 50\% of the total reported incidents it receives \citep{weisburd2015law}. This approach is mostly guided by predictive techniques that forecast crime in both time and space \citep{berk2021artificial}, while the response may involve strategies such as increasing foot patrols, engaging with the local stakeholders, or implementing problem-oriented policing \citep{braga2008policing, taylor2011randomized, weisburd2022reforming}.

The main issue with the above perspective is that an over-reliance on purely predictive methods fails to inform on the cascading effects of {\em hot spot} interventions,  which include the risk of crime and disorder displacement. Critics of intensive {\em hot spot} policing argue that such strategies tend to displace crime rather than effectively reduce it. The literature is inconclusive and cannot rule out positive or negative spillover effects. \citeauthor{braga2019hot}'s (\citeyear{braga2019hot}) review shows that studies usually do not detect noticeable negative spillover to other areas. Several analyses even identify benefits that spread from {\em hot spots} to adjacent areas \citep[e.g.][]{weisburd2006does}. Both incapacitation and deterrence can explain these results. If the arrested offenders also operated in neighboring areas, the offense is expected to decrease in the {\em hot spots}  and the surrounding areas. On the other hand, some studies document crime displacement when the police strategy focuses on specific locations. For example, a study by \citet{blattman2017state} finds that any crime deterrence in the target area is more than offset by increased crime on the streets within the following 250 meters. In addition, such an analysis unveils heterogeneous effects by crime category. While violent crimes decrease, property crimes increase. The authors conclude that this type of policing pushes property crime ``around the corner''. Similarly, \cite{banerjee2019efficient} use randomized drunk driving checkpoints to find  strategic responses that explain the displacement of crime to other areas.

The lack of a consensus in the aforementioned studies underscores the need for a more detailed understanding of how various crime categories  manifest, and interact, within different urban areas across time. Even though crime clustering in time and space is a known phenomenon, an unanimous consensus remains elusive regarding the specific interactions among criminal activities in time, space and across crime categories. Consequently, an ongoing debate surrounds the tangible benefits of specific intervention policies, particularly those with a local emphasis (such as the  {\em hot spots} approach) based on black-box predictive methods. In fact, these methods are not designed to provide a model-based and interpretable understanding of complex dependencies in three-way criminal activity tensors. In Section~\ref{sec:model}, we cover this gap through a statistical model that can disentangle these dependencies, while preserving a predictive performance competitive with routine use black-box methods. As such, the proposed perspective can open the avenues for refining the design and evaluation of law enforcement interventions by combining accurate forecasts with a detailed quantitative assessment of the spillover effects that these interventions may trigger. To the best of our knowledge, this direction has received limited attention in the literature.

\subsection{\large 2.3. Existing statistical approaches}
\label{mod}
As discussed in Section~\ref{sec_1},  available model-based solutions for the study of criminal activity data \citep[][]{balocchi2023crime, balocchi2019spatial,clark2018modeling,law2014bayesian, liu2017joint, quick2018crime, shoesmith2013space,vicente2023multivariate} mostly constrain the analysis to two selected dimensions of the observed three-way tensor. Although such a perspective fails to characterize and learn the full range of dependencies across time, space and crime categories, the available contributions along these directions highlight several important aspects that are the basis of the statistical model we design in Section~\ref{sec:model}.

More specifically, in line with the discussion in Sections~\ref{det}--\ref{pol}, the design of model-based representations for the complex dependencies across time and space is crucial in analyzing criminal activity data. For example, \citet{shoesmith2013space} demonstrates the importance of temporal lags and spatio-temporal interactions when modeling criminal activities. Similarly, \citet{balocchi2019spatial} and \citet{balocchi2023crime} clarify how  borrowing of information among the temporal crime trends across space improves modeling outcomes. \citet{vicente2023multivariate} employ, instead, a penalized spline expansion across locations and times, further allowing for joint modeling of different crime categories. Although such a formulation does not explicitly incorporate graphical dependencies among these different crime categories, it provides relevant support to the importance of devising joint formulations for the entire three-way criminal activity tensor. This support can be found also in the static spatial models by, e.g. \citet{liu2017joint} and  \citet{quick2018crime}, which highlight how a multivariate perspective focusing on multiple crime categories can unveil structural patterns that a univariate model is unable to detect. While black-box strategies \citep[e.g.][]{berk2021artificial,mandalapu2023crime}  can jointly leverage the complex system of dependencies within the observed three-way criminal activity tensor to enhance predictive accuracy, the design of a statistical model that can disentangle and infer  these dependencies, without sacrificing forecasting performance, is substantially more challenging. Although such a perspective has received limited attention, as discussed in Section~\ref{pol}, advancements along this direction are necessary to devise strategies based not only on accurate forecasts, but also on a careful assessment of the cascading effects of the adopted intervention strategy. The relevance of this endeavor has motivated some efforts towards providing structural interpretation to black-box algorithms in crime analysis, but these interpretations are based on ex-post variable importance measures when external covariates are related to crime activity via a simple regression structure \citep[e.g.][]{zhang2022interpretable}.

In Section~\ref{sec:model}, we cover such a gap via a Bayesian model-based formulation that balances parsimony, interpretability and predictive accuracy by integrating ideas from matrix autoregressive processes, bilinear constructions, low-rank factorizations and graphical models.

\section{\large 3. Low-rank bilinear autoregressive model for criminal activities}
\label{sec:model}
Following similar practice in the analysis of criminal activity data \citep[see, e.g.][]{balocchi2019spatial,shoesmith2013space}, we model the counts $c_{kqt} \in \mathbb{N}$ of the reported crimes for every category $k = 1, \dots, K$, in location $q = 1, \dots, Q$ at time $t = 1, \dots, T$ under the $\log(1+c_{kqt})$ transformation. Although modeling the original counts is also a possibility, such an option would require Poisson or negative Binomial likelihoods, which, however, have witnessed less extensions and methodological advancements in tensor-valued data settings compared to the classical Gaussian framework employed in practice for  $\log(1+c_{kqt})$ \citep[][]{balocchi2019spatial,shoesmith2013space}. 

As illustrated in Section~\ref{sec:data} (see Figure~\ref{fig:log_counts}), the log-transformed counts often exhibit systematic  time trends and seasonal components, for several combinations of crime categories and spatial locations. These trends and seasonal effects  might obscure the system of stochastic dependencies that are the focus of our analysis. Following standard practice \citep[e.g.][]{ham_1994,tsay2005analysis}, we thus pre-process the data by removing the quadratic trends and seasonal components displayed by the log-transformed series in Figure~\ref{fig:log_counts}. This is accomplished by subtracting from $\log(1+c_{kqt})$ the term $\hat{\mu}_{kqt}=\hat{\beta}_{0kq}+\hat{\beta}_{1kq}t+\hat{\beta}_{2kq}t^2+\hat{\boldsymbol \gamma}^{\intercal}_{kq}{\bf d}_t$, for each $k = 1, \dots, K$, $q = 1, \dots, Q$ and $t=1, \ldots, T$, where $\hat{\boldsymbol \gamma}_{kq}$ comprises the monthly seasonal effects, while ${\bf d}_t$ is a vector of dummy variables which identify the month associated with the time index $t$. The estimates of these coefficients can be readily obtained via \textsc{ols} applied separately, for every pair $(k,q)$, to the regression model having response $\log(1+c_{kqt})$ and predictor ${\mu}_{kqt}$, for $t=1, \ldots, T$ (refer to Figure~B1 in the Supplementary Material for a graphical representation of the resulting pre-processed time series). Notice that an alternative to this series-by-series detrending and seasonal adjustment would be to specify a joint model for the trend and seasonal components across crimes and districts, which facilitates borrowing of information in estimating the associated parameters. Although such a perspective provides an interesting direction for future research, our two-stage pipeline---i.e. first removing systematic trends and seasonal patterns with a simple adjustment strategy and then designing a more structured model to learn complex dependencies within the adjusted series---is consistent with routine practice \citep[e.g.][]{ham_1994,tsay2005analysis,vatter2015generalized,zhang2005neural} and facilitates the overall analysis. 

Moreover, as discussed in Section~\ref{predictive}, the two-stage pipeline we employ achieves a predictive performance comparable to those of  powerful black-box strategies applied directly to the observed log-counts (without detrending and seasonal adjustment), thereby confirming its effectiveness. Note that, under our two-stage perspective, the predicted crime levels on the original scale can be obtained by adding  $\hat{\mu}_{kqt}$ to the forecasts of the model designed in Section~\ref{sec:formul} for the detrended and seasonally-adjusted log-transformed counts, which we denote as $y_{kqt}$. More specifically, let $\bf{Y}$ be the $K \times Q \times T$ adjusted criminal activity tensor having entries $y_{kqt}$, for each $k = 1, \dots, K$, $q = 1, \dots, Q$ and $t = 1, \dots, T$. In Section~\ref{sec:formul}, we design a statistical model for $\bf{Y}$ that characterizes the temporal dependence among its $K \times Q$ (crime-category -- location) slices ${\bf Y}_t$, at each time index $t = 1, \dots, T$, via a matrix autoregressive construction separating spatial and crime-category effects through a  low-rank bilinear formulation. This model, combined with a separable covariance structure for the error term, allows to infer spatial dependencies, interlinkages between crime categories and persistence over time, without undermining predictive accuracy. Inference under the proposed model proceeds via a Bayesian framework outlined in Section~\ref{sec_3}.

\subsection{\large 3.1. Model formulation}
\label{sec:formul}
To devise a model that captures spatial, temporal and crime-category effects within the three-way criminal activity tensor, while facilitating the design of  forecasting strategies, a simple building-block option is to rely on a full autoregressive construction for the matrix-valued sequence ${\bf Y}_t$, $t = 1, \dots, T$. More specifically, let ${\bf y}_t= \mbox{vec}({\bf Y}_t)=[y_{11t}, \dots, y_{kqt}, \dots y_{KQt}]^{\intercal}$ be the vectorized version of the matrix ${\bf Y}_t$, for each time $t = 1, \dots, T$, then it is natural to start by assuming that ${\bf y}_{t} = \bm{\Theta}{\bf y}_{t-1} + \bm{\varepsilon}_t $, with $\bm{\varepsilon}_t \sim \mbox{N}_{KQ}({\bf 0}, \bm{\Sigma})$ independently for every  $t = 2, \dots, T$. In this standard vector autoregressive formulation the lagged values within ${\bf y}_{t-1}$ affect the current ones in ${\bf y}_{t}$ via a full and unstructured $KQ \times KQ$ coefficient matrix $\bm{\Theta}$, which captures the different inter-temporal effects that every combination of location and crime category exerts on each entry in ${\bf y}_{t}$, whereas the $KQ \times KQ$ covariance matrix $\bm{\Sigma}$ encodes instantaneous dependencies in reaction to shocks. 

Although unstructured vector autoregressive constructions are not novel in modeling of criminal activity data \citep[e.g.][]{funk2003identifying,shoesmith2013space}, none of the available solutions is designed to study  time changes of criminal activities jointly across space and crime categories. In fact, current solutions focus  only on one of the two dimensions. As discussed in Section~\ref{sec:lit}, this choice is practically and conceptually suboptimal, since there are nuanced interactions among types of crime and spatial locations across time. Accounting explicitly for both, as in the above model, facilitates borrowing of information to improve forecasts and opens the avenues to provide interpretable inference on these interactions. To accomplish this goal it is however important to consider a structured representation for both the coefficient matrix $\bm{\Theta}$ and the error covariance matrix $\bm{\Sigma}$ in order to disentangle spatial and crime-category dependencies, while avoiding overparameterized formulations with potential overfitting issues. 

Motivated by the above considerations, we thus rely on a separable construction that assumes $\bm{\Theta}={\bf B}\otimes {\bf A}$ and $\bm{\Sigma}=\bm{\Sigma_{\bf B}} \otimes \bm{\Sigma_{\bf A}}$, where $\otimes$ denotes the Kronecker product.  In this factorization, ${\bf A}$ and ${\bf B}$  are $K \times K$ and  $Q \times Q$ matrices that capture, respectively, the inter-temporal dependence between the $K$ crime categories shared across the spatial locations, and among these $Q$ spatial locations within each crime category. $\mathbf A$ is common across all districts and $\mathbf B$ is common across all crime categories, yielding shared dependence patterns along the two modes. Similarly, $\bm{\Sigma}_{ \bf A}$ and $\bm{\Sigma}_{ \bf B}$ correspond to the  $K \times K$ and  $Q \times Q$ covariance matrices encoding the across-categories and across-locations instantaneous dependencies in reaction to shocks, respectively. In this way, it is possible to disentangle spatial and crime-category dependencies, thereby enhancing interpretability, while preserving flexibility by letting these dependencies interact in the definition of $\bm{\Theta}$ and $\bm{\Sigma}$. Separable constructions have been successfully employed in black-box methods \citep[e.g.][]{wang2025mragnn} to improve predictive accuracy, but the potential of these perspectives to allow for interpretable model-based inference on the complex dependencies within the criminal activity tensor have not yet  been explored. Notice that, since we remove trend and seasonal components prior to modeling these complex dependencies, the matrices ${\bf A}$, ${\bf B}$, $\bm{\Sigma}_{\bf A}$ and $\bm{\Sigma}_{\bf B}$ characterize statistical associations in the deviations from each series' systematic trend and seasonal patterns. As discussed previously, this perspective is common in the analysis of temporal data. 

Leveraging properties of the Kronecker product and of matrix-variate normal  (MN) distributions, the above separable autoregressive construction admits the following bilinear representation
\begin{eqnarray}
\label{eq:main_eq}
    {\bf Y}_t = {\bf A}{\bf Y}_{t-1}{\bf B}^{\intercal} + {\bf E}_t, \qquad {\bf E}_t \sim \mbox{MN}_{K,Q}({\bf 0}, \bm{\Sigma}_{\bf A},\bm{\Sigma}_{\bf B}),
\end{eqnarray}
\noindent independently for each $t=2, \ldots, T$, where ${\bf E}_t$ is the  $K \times Q$ matrix with noise entries $\varepsilon_{kqt}$, while ${\bf E}_t \sim \mbox{MN}_{K,Q}({\bf 0}, \bm{\Sigma}_{\bf A},\bm{\Sigma}_{\bf B})$ implies $\mbox{vec}( {\bf E}_t)=\bm{\varepsilon}_t \sim \mbox{N}_{KQ}({\bf 0},\bm{\Sigma_{\bf B}} \otimes \bm{\Sigma_{\bf A}})$. Therefore, under~\eqref{eq:main_eq}, the model for each entry $y_{kqt}$ of ${\bf Y}_t$ is 
\begin{eqnarray*}
y_{kqt}=\sum_{k'=1}^K\sum_{q'=1}^Q a_{kk'}b_{qq'}y_{k'q'(t-1)}+\varepsilon_{kqt},
\end{eqnarray*}
where  $a_{kk'}$ measures the influence of crime category $k'$ on the $k$--th one, shared across districts, while $b_{q q'}$ quantifies the global effect of district $q$ on $q'$, common to crime categories. As such, changes in $y_{k' q'(t-1)}$ will propagate, positively or negatively, to crime categories with high $|a_{kk'}|$, within districts having large $|b_{qq'}|$.

The above formulation specializes the class of multilinear tensor regression models proposed by \cite{hoff2015multilinear}\footnote{Note that while the original focus of  \cite{hoff2015multilinear} is on dynamic networks where ${\bf Y}_t$ is an adjacency matrix encoding connectivity data, the formulation proposed extends to general tensor-valued data, beyond dynamic networks. In our setting ${\bf Y}_t$ does not measure directed relationships, but rather crime levels at time $t$, for each combination of crime category $k$ and spatial location $q$, denoting, respectively, the rows and columns of ${\bf Y}_t$. As such, the ordering within rows and columns does not play a role in the model formulation and results.} to our  setting by relating a matrix-valued outcome to a matrix-variate input via a temporal autoregressive structure; see \citet{billio2023bayesian}, \citet{martinez2016towards} and the references therein for general multi-way tensor regression constructions with focus on disease mapping  and econometric applications, respectively. Crucially, compared to a fully unstructured vector autoregressive formulation, the model within \eqref{eq:main_eq} reduces the number of parameters from $(KQ)^2 + (KQ)(KQ+1)/2$ to  $K^2 + Q^2 + K(K+1)/2 + Q(Q+1)/2$.  In our application to Chicago criminal activity data with $K=28$ crime categories and $Q=22$ spatial locations, this implies a reduction from $569{,}492$ parameters to $1{,}927$  (about three hundred times lower than those within the unstructured model). Interestingly, as illustrated in Section~\ref{sec:application}, such a remarkable reduction in complexity still yields a predictive performance competitive with routine use black-box methods, meaning that the model in \eqref{eq:main_eq} properly accounts for a rich set of dynamics within and between crime categories, locations and times.

Although the above formulation achieves a sensible balance between predictive potential and interpretable inference on the dependencies in the criminal activity tensor, in practice, unrestricted estimation of the \(K \times K\) and \(Q \times Q\) matrices ${\bf A}$ and ${\bf B}$ may still result in a overly-parametrized model. In addition, as discussed in Section~\ref{sec:lit}, it is reasonable to expect that both ${\bf A}$ and ${\bf B}$ admit lower-dimensional representations due to the sparse and structured inter-temporal dependencies associated with both crime categories and spatial locations. To account for this information and further reduce complexity, we combine the model in \eqref{eq:main_eq} with a low-rank factorization for ${\bf A}$ and ${\bf B}$ defined as
\begin{eqnarray}
\label{eq:fact}
   {\bf A} = {\bf L}_{\bf A}{\bf Z}_{\bf A}, \qquad  \qquad  {\bf B} = {\bf L}_{\bf B}{\bf Z}_{\bf B}, 
\end{eqnarray}
 where ${\bf L}_{\bf A}$ and ${\bf L}_{\bf B}$ have dimension \(K \times R_{\bf A}\) and \(Q \times R_{\bf B}\), respectively, with $R_{\bf A} \leq K$ and $R_{\bf B} \leq Q$. Conversely, the matrices ${\bf Z}_{\bf A}$ and ${\bf Z}_{\bf B}$ are of dimension \(R_{\bf A} \times K\) and \(R_{\bf B} \times Q\). In this way, the fixed integers \(R_{\bf A}\) and \(R_{\bf B}\) determine the rank of the coefficient matrices ${\bf A}$ and ${\bf B}$, thereby further reducing the number of free parameters in both  ${\bf A}$ and ${\bf B}$. This allows for additional reductions in complexity compared to a full-rank matrix-autoregressive model. In the motivating application,  \(R_{\bf A}\) and \(R_{\bf B}\) are specified based on the results of an out-of-sample predictive exercise. See \citet{gregory2022multivariate} and \citet{tsay2024matrix} for related factorizations in reduced-rank multivariate regression models and for recent developments in matrix-variate autoregressive settings.

Despite the potential of the above ideas for learning structure within three-way criminal activity tensors, to our knowledge formulations of this type have been underutilized in criminology. To cover such a gap we employ the model in \eqref{eq:main_eq}--\eqref{eq:fact} and perform inference on its parameters under a Bayesian perspective (see Section~\ref{sec_3}). This choice provides  a fully probabilistic framework enabling formal uncertainty quantification and possible inclusion of expert information along with nuanced hierarchical structures for more sophisticated crime modeling exercises.

\section{\large 4. Bayesian inference and computation}\label{sec_3}

\subsection{\large 4.1. Prior specification}
\label{prior}
To perform Bayesian inference under model  \eqref{eq:main_eq}--\eqref{eq:fact} it is necessary to elicit prior distributions for its parameters $\bm{\Sigma}_{\bf A}$, $\bm{\Sigma}_{\bf B}$, ${\bf L}_{\bf A}$, ${\bf L}_{\bf B}$, ${\bf Z}_{\bf A}$ and ${\bf Z}_{\bf B}$. These priors are specified to facilitate tractable posterior inference, while favoring weakly-informative or non-informative elicitations that do not overwhelm the information provided by the observed  tensor via the likelihood of model~\eqref{eq:main_eq}--\eqref{eq:fact}.
 
Consistent with the above goals, we follow routine practice in Bayesian modeling of covariance matrices \citep[e.g.][]{gelman1995bayesian} by relying on inverse-Wishart priors for  $\bm{\Sigma}_{\bf A}$ and $\bm{\Sigma}_{\bf B}$; namely
\begin{eqnarray*}
\bm{\Sigma}_{\bf A} \sim \mbox{inv-Wishart}(2, 2\cdot{\bf I}_K), \qquad \quad \bm{\Sigma}_{\bf B} \sim \mbox{inv-Wishart}(2, 2\cdot{\bf I}_Q),
\end{eqnarray*}
where the hyperparameters are selected to induce an improper specification that proved effective both in inference and in out-of-sample prediction for the application in Section~\ref{sec:application} (when tested in sensitivity analyses, moderate changes of these  hyperparameters did not substantially affect the final results). Combined with the Gaussian likelihood induced by model \eqref{eq:main_eq}--\eqref{eq:fact}, the above priors admit conditional-conjugacy properties, which facilitate Bayesian inference on the instantaneous dependence structures among the $K$ crime categories and between the $Q$ spatial locations. Note that these priors do not include specific structures within $\bm{\Sigma}_{\bf A}$ and $\bm{\Sigma}_{\bf B}$, but rather learn flexibly both matrices from the observed data. This choice may appear suboptimal for $\bm{\Sigma}_{\bf B}$, which measures dependence across locations, thereby suggesting alternative representations that explicitly include spatial proximity information regulating these dependencies, such as \textsc{car}-type constructions \citep[][]{banerjee2003hierarchical}. Although it would be interesting to extend the proposed framework to explicitly incorporate such information, this perspective would undermine the flexibility of the model in our setting. In fact,  as anticipated in Section~\ref{sec:lit}, the spatial dependencies in the criminal activities typically arise from the combined effect of several underlying factors such as localized presence of criminal organizations or distinctive policing strategies, which are difficult to include in the model only via spatial proximity information. Avoiding such constraints allows us to learn in Section~\ref{sec:application} (see Figures~\ref{fig:networks}--\ref{fig:district_communities}) dependencies among Chicago districts that partially depart from pure spatial proximity and provide quantitative evidence supporting the reported presence of organized crime and gangs activities in the city \citep[e.g.][]{aspholm2019views,schnell2017influence}. 

As for the priors on ${\bf Z}_{\bf A}$ and ${\bf Z}_{\bf B}$, we adapt to our  low-rank bilinear autoregressive setting the elicitation suggested for the general class of multilinear tensor regression models in \citet{hoff2015multilinear}. In particular, we assume independent $\mbox{N}(0,10)$ priors for all the entries of ${\bf Z}_{\bf A}$ and ${\bf Z}_{\bf B}$, where the variance of $10$ yields a weakly-diffuse specification that introduces some regularization effect, but avoids imposing excessive information through the prior. In matrix form, this choice implies
\begin{eqnarray*}
{\bf Z}_{\bf A} \sim  \mbox{MN}_{R_{\bf A},K}({\bf 0}, {\bf I}_{R_{\bf A}},10 \cdot {\bf I}_K), \qquad \quad {\bf Z}_{\bf B} \sim \mbox{MN}_{R_{\bf B},Q}({\bf 0}, {\bf I}_{R_{\bf B}},10\cdot {\bf I}_Q),
\end{eqnarray*}
thereby inducing via the low-rank construction in \eqref{eq:fact} weakly-diffuse and conditionally-conjugate  priors also on the separable inter-temporal effects in ${\bf A}$ and ${\bf B}$. Under \eqref{eq:fact}, such effects are also controlled by ${\bf L}_{\bf A}$ and ${\bf L}_{\bf B}$. Consistent with the previous elicitations, the entries of such matrices are assigned non-informative flat priors (i.e.\ uniform in $\mathbb{R}$) further combined with a processing step in the proposed Gibbs sampler that enforces the constraints  ${\bf L}_{\bf A}^{\intercal} {\bf L}_{\bf A}= {\bf I}_{R_{\bf A}}$ and ${\bf L}_{\bf B}^{\intercal} {\bf L}_{\bf B}= {\bf I}_{R_{\bf B}}$, respectively. This choice implicitly fixes the scale of the elements within ${\bf L}_{\bf A}$ and ${\bf L}_{\bf B}$, thereby addressing rotational and scaling identification issues that may arise from the factorization in \eqref{eq:fact}. While these identifiability problems only affect ${\bf L}_{\bf A}$, ${\bf Z}_{\bf A}$ and ${\bf L}_{\bf  B}$, ${\bf Z}_{\bf  B}$, and not the main objects of inference, i.e.\ ${\bf A}$ and ${\bf B}$, considering processing steps that address these potential issues improves mixing and convergence of the Gibbs sampler we design in Section~\ref{sec:gibbs} for posterior inference.

\subsection{\large 4.2 Posterior computation via Gibbs sampling} \label{sec:gibbs}
The priors  in Section~\ref{prior} combined with the likelihood  of model \eqref{eq:main_eq}--\eqref{eq:fact} yield a posterior for the parameters $\bm{\Sigma}_{\bf A}$, $\bm{\Sigma}_{\bf B}$, ${\bf L}_{\bf A}$, ${\bf L}_{\bf B}$, ${\bf Z}_{\bf A}$ and ${\bf Z}_{\bf B}$ which is not available in closed form. Nonetheless, tractable full conditionals for each of these parameters, given the others and the observed data are available, thereby allowing for Monte Carlo inference leveraging posterior samples  of these parameters obtained via a Gibbs sampling algorithm.

To derive in detail the different steps of such a sampling algorithm, let  ${{\bf Y}}_{{\bf A}}=[{\bf Y}_{2}, \ldots, {\bf Y}_{T}]$ and  ${{\bf Y}}_{{\bf B}}=[{\bf Y}^{\intercal}_{2}, \ldots, {\bf Y}^{\intercal}_{T}]$. In addition, denote with ${{\bf X}}_{{\bf A}}$ and ${{\bf X}}_{{\bf B}}$ the lagged versions of ${{\bf Y}}_{{\bf A}}$ and ${{\bf Y}}_{{\bf B}}$ respectively. Then, under these settings, and conditionally on ${\bf B}$ and $\bm{\Sigma}_{\bf B}$, model \eqref{eq:main_eq} can be expressed jointly over the whole time window as
\begin{eqnarray}
\label{eq:a_matrix}
    \Tilde{{\bf Y}}_{\bf A} = {\bf A}{\Tilde{\bf X}}_{\bf A} + \Tilde{{\bf E}}_{\bf A}, \quad \quad  \Tilde{{\bf E}}_{\bf A} \sim  \mbox{MN}_{K,(T-1)Q}({\bf 0}, \bm{\Sigma}_{\bf A},{\bf I}_{(T-1)Q}),
\end{eqnarray}
with $\Tilde{{\bf Y}}_{\bf A} = {{\bf Y}}_{\bf A}({\bf I}_{T-1}\otimes \bm{\Sigma}_{\bf B}^{-1/2})$ and ${\Tilde{\bf X}}_{\bf A} = {{\bf X}}_{\bf A}({\bf I}_{T-1}\otimes {\bf B}^{\intercal}\bm{\Sigma}_{\bf B}^{-1/2})$, where $\bm{\Sigma}_{\bf B}^{-1/2}$ is the Cholesky decomposition of the inverse of $\bm{\Sigma}_{\bf B}$. We refer to \eqref{eq:a_matrix} as the  {\em $A$-matricization} of (\ref{eq:main_eq}). Leveraging the low-rank factorization \eqref{eq:fact}  of ${\bf A}$ and basic properties of vectorizations, we can re-express~\eqref{eq:a_matrix} as
\begin{eqnarray}
\label{eq:Z_fac}
    \text{vec}(\Tilde{{\bf Y}}_{\bf A}) = ({\Tilde{\bf X}}_{\bf A}^{\intercal} \otimes {\bf L}_{\bf A}) \text{vec}({\bf Z}_{\bf A}) + \text{vec}(\Tilde{{\bf E}}_{\bf A}), \ \ \text{vec}(\Tilde{{\bf E}}_{\bf A}) \sim \mbox{N}_{K(T-1)Q}({\bf 0}, {\bf I}_{(T-1)Q} \otimes \bm{\Sigma}_{\bf A}),
\end{eqnarray}
or, equivalently, via 
\begin{eqnarray}
\label{eq:L_fac}
    \text{vec}(\Tilde{{\bf Y}}_{\bf A}) =  ({\Tilde{\bf X}}_{\bf A}^{\intercal} {\bf Z}_{\bf A}^{\intercal} \otimes {\bf I}_K) \text{vec}({\bf L}_{\bf A}) + \text{vec}(\Tilde{{\bf E}}_{\bf A}),  \ \ \text{vec}(\Tilde{{\bf E}}_{\bf A}) \sim \mbox{N}_{K(T-1)Q}({\bf 0}, {\bf I}_{(T-1)Q} \otimes \bm{\Sigma}_{\bf A}).
\end{eqnarray}
This allows to recast the model in \eqref{eq:main_eq}--\eqref{eq:fact} as a Gaussian linear regression with coefficients $\text{vec}({\bf Z}_{\bf A})$ (see~\eqref{eq:Z_fac}) or $\text{vec}({\bf L}_{\bf A})$ (see~\eqref{eq:L_fac}), respectively, conditionally on the other parameters. Combined with the priors for $\text{vec}({\bf Z}_{\bf A})$ and $\text{vec}({\bf L}_{\bf A})$ presented in Section~\ref{prior}, such a formulation yields Gaussian full conditionals for both vectors. More specifically, $(\text{vec}({\bf Z}_{\bf A}) \mid -) \sim \mbox{N}_{R_{\bf A} K}(\bm{\mu}_{{\bf Z}_{\bf A}},\bm{\Sigma}_{{\bf Z}_{\bf A}})$ with
    \begin{eqnarray}
    \label{post_Z_A}
    \begin{split}
    \bm{\Sigma}_{{\bf Z}_{\bf A}} &= [10^{-1}{\bf I}_{R_{\bf A}K} + ({\Tilde{\bf X}}_{\bf A}^{\intercal} \otimes {\bf L}_{\bf A})^{\intercal} ({\bf I}_{(T-1)Q} \otimes \bm{\Sigma}_{\bf A})^{-1}({\Tilde{\bf X}}_{\bf A}^{\intercal} \otimes {\bf L}_{\bf A})]^{-1}\\
    &= [10^{-1}{\bf I}_{R_{\bf A}K} + ({\Tilde{\bf X}}_{\bf A}{\Tilde{\bf X}}_{\bf A}^{\intercal} \otimes {\bf L}_{\bf A}^{\intercal}\bm{\Sigma}_{\bf A}^{-1}{\bf L}_{\bf A})]^{-1},\\
  \bm{\mu}_{{\bf Z}_{\bf A}} &=  \bm{\Sigma}_{{\bf Z}_{\bf A}}  ({\Tilde{\bf X}}_{\bf A}^{\intercal} \otimes {\bf L}_{\bf A})^{\intercal} ({\bf I}_{(T-1)Q} \otimes \bm{\Sigma}_{\bf A})^{-1} \text{vec}(\Tilde{{\bf Y}}_{\bf A})= \bm{\Sigma}_{{\bf Z}_{\bf A}}      \text{vec}({\bf L}_{\bf A}^{\intercal}\bm{\Sigma}_{\bf A}^{-1}\Tilde{{\bf Y}}_{\bf A}{\Tilde{\bf X}}_{\bf A}^{\intercal}).
    \end{split}
    \end{eqnarray}
Similarly,     $(\text{vec}({\bf L}_{\bf A}) \mid -) \sim \mbox{N}_{R_{\bf A} K}(\bm{\mu}_{{\bf L}_{\bf A}},\bm{\Sigma}_{{\bf L}_{\bf A}})$ with $\bm{\mu}_{{\bf L}_{\bf A}}$ and $\bm{\Sigma}_{{\bf L}_{\bf A}}$ defined as
      \begin{eqnarray}
    \label{post_L_A}
    \begin{split}
    \bm{\Sigma}_{{\bf L}_{\bf A}} &= [({\Tilde{\bf X}}_{\bf A}^{\intercal}{\bf Z}_{\bf A}^{\intercal} \otimes {\bf I}_K)^{\intercal} ({\bf I}_{(T-1)Q} \otimes \bm{\Sigma}_{\bf A})^{-1}({\Tilde{\bf X}}_{\bf A}^{\intercal}{\bf Z}_{\bf A}^{\intercal} \otimes {\bf I}_K)]^{-1}\\
    &= ({\bf Z}_{\bf A}{\Tilde{\bf X}}_{\bf A}{\Tilde{\bf X}}_{\bf A}^{\intercal}{\bf Z}_{\bf A}^{\intercal} \otimes \bm{\Sigma}_{\bf A}^{-1})^{-1},\\
  \bm{\mu}_{{\bf L}_{\bf A}} &=   \bm{\Sigma}_{{\bf L}_{\bf A}} ({\Tilde{\bf X}}_{\bf A}^{\intercal}{\bf Z}_{\bf A}^{\intercal} \otimes {\bf I}_K)^{\intercal} ({\bf I}_{(T-1)Q} \otimes \bm{\Sigma}_{\bf A})^{-1} \text{vec}(\Tilde{{\bf Y}}_{\bf A})= \bm{\Sigma}_{{\bf L}_{\bf A}}  \text{vec}(\bm{\Sigma}_{\bf A}^{-1}\Tilde{{\bf Y}}_{\bf A}{\Tilde{\bf X}}_{\bf A}^{\intercal} {\bf Z}_{\bf A}^{\intercal}).
    \end{split}
    \end{eqnarray}
Leveraging the above samples for $\text{vec}({\bf Z}_{\bf A}) $ and $\text{vec}({\bf L}_{\bf A})$, a draw from the full conditional  of ${\bf A}$ can be obtained by computing  ${\bf A} = {\bf L}_{\bf A}{\bf Z}_{\bf A}$ at these samples. Conditioned on ${\bf A}$, it is then possible to leverage the conjugacy of the inverse-Wishart prior for $\bm{\Sigma}_{\bf A}$ under  \eqref{eq:a_matrix}, to obtain
\begin{eqnarray}
(\bm{\Sigma}_{\bf A} \mid -) \sim \mbox{inv-Wishart}(2 + (T-1)Q, 2\cdot{\bf I}_K +  ( \Tilde{{\bf Y}}_{\bf A} - {\bf A}{\Tilde{\bf X}}_{\bf A})( \Tilde{{\bf Y}}_{\bf A} - {\bf A}{\Tilde{\bf X}}_{\bf A})^{\intercal}).
    \label{post_Sigma_A}
\end{eqnarray}
    
A similar reasoning and derivations apply to the full conditionals for    ${\bf Z}_{\bf B}$, ${\bf L}_{\bf B}$ and $\bm{\Sigma}_{\bf B}$. More specifically, given ${\bf A}$ and  $\bm{\Sigma}_{\bf A}$, it is possible to obtain the following  {\em $B$-matricization} of (\ref{eq:main_eq})
\begin{eqnarray}
\label{eq:b_matrix}
    \Tilde{{\bf Y}}_{\bf B} = {\bf B}{\Tilde{\bf X}}_{\bf B} + \Tilde{{\bf E}}_{\bf B} \quad \quad  \Tilde{{\bf E}}_{\bf B} \sim \mbox{MN}_{Q,(T-1)K}({\bf 0}, \bm{\Sigma}_{\bf B}, {\bf I}_{(T-1)K})
\end{eqnarray}
where $\Tilde{{\bf Y}}_{\bf B} = {{\bf Y}}_{\bf B}({\bf I}_{T-1}\otimes\bm{\Sigma}_{\bf A}^{-1/2})$ and ${\Tilde{\bf X}}_{\bf B} = {{\bf X}}_{\bf B}({\bf I}_{T-1}\otimes{\bf A}^{\intercal}\bm{\Sigma}_{\bf A}^{-1/2})$, with $\bm{\Sigma}_{\bf A}^{-1/2}$ denoting the Cholesky decomposition of the inverse of $\bm{\Sigma}_A$. Hence, as in \eqref{eq:Z_fac}--\eqref{eq:L_fac}, also in this case it is possible to express~\eqref{eq:b_matrix} in the form of a conditionally-Gaussian linear regression 
\begin{eqnarray}
\label{eq:Z_fac_B}
    \text{vec}(\Tilde{{\bf Y}}_{\bf B}) = ({\Tilde{\bf X}}_{\bf B}^{\intercal} \otimes {\bf L}_{\bf B}) \text{vec}({\bf Z}_{\bf B}) + \text{vec}(\Tilde{{\bf E}}_{\bf B}), \ \ \text{vec}(\Tilde{{\bf E}}_{\bf B}) \sim \mbox{N}_{Q(T-1)K}({\bf 0}, {\bf I}_{(T-1)K} \otimes \bm{\Sigma}_{\bf B}),
\end{eqnarray}
with coefficients $\text{vec}({\bf Z}_{\bf B}) $, or via the alternative formulation
\begin{eqnarray}
\label{eq:L_fac_B}
    \text{vec}(\Tilde{{\bf Y}}_{\bf B}) =  ({\Tilde{\bf X}}_{\bf B}^{\intercal} {\bf Z}_{\bf B}^{\intercal} \otimes {\bf I}_Q) \text{vec}({\bf L}_{\bf B}) + \text{vec}(\Tilde{{\bf E}}_{\bf B}), \ \ \text{vec}(\Tilde{{\bf E}}_{\bf B}) \sim \mbox{N}_{Q(T-1)K}({\bf 0}, {\bf I}_{(T-1)K} \otimes \bm{\Sigma}_{\bf B}),
\end{eqnarray}
parameterized by $\text{vec}({\bf L}_{\bf B})$.

 \begin{algorithm}[t]
\caption{Gibbs sampler for model  \eqref{eq:main_eq}--\eqref{eq:fact}}\label{alg:gibbs}
\begin{algorithmic}
\For{each Gibbs iteration}
    \State 1. sample $\text{vec}({\bf Z}_{\bf A})$  from the Gaussian full conditional with parameters as in \eqref{post_Z_A}
    \State 2. sample $\text{vec}({\bf L}_{\bf A})$  from the Gaussian full conditional with parameters as in \eqref{post_L_A}
        \State 3. transform the sampled ${\bf Z}_{\bf A}$ and ${\bf L}_{\bf A}$  to enforce the constraint discussed in Section~\ref{prior}
\State 4. sample $\bm{\Sigma}_{\bf A}$ from the inverse-Wishart full conditional  in \eqref{post_Sigma_A} 
    \State 5. sample $\text{vec}({\bf Z}_{\bf B})$  from the Gaussian full conditional with parameters as in \eqref{post_Z_B}
    \State 6. sample $\text{vec}({\bf L}_{\bf B})$  from the Gaussian full conditional with parameters as in \eqref{post_L_B}
      \State 7. transform the sampled ${\bf Z}_{\bf B}$ and ${\bf L}_{\bf B}$  to enforce the constraint discussed in Section~\ref{prior}
\State 8. sample $\bm{\Sigma}_{\bf B}$ from the inverse-Wishart full conditional  in \eqref{post_Sigma_B} 
\State 9. scale $\bm{\Sigma}_{\bf A}$, $\bm{\Sigma}_{\bf B}$, ${\bf A}= \mathbf{L}_{\bf A} \mathbf{Z}_{\bf A}$ and ${\bf B}=\mathbf{L}_{\bf B} \mathbf{Z}_{\bf B}$ to ensure identifiability
\EndFor
\end{algorithmic}
\end{algorithm}

Combining \eqref{eq:Z_fac_B} with the conditionally conjugate matrix-variate normal prior for ${\bf Z}_{\bf B}$, implies that $(\text{vec}({\bf Z}_{\bf B}) \mid -) \sim \mbox{N}_{R_{\bf B} Q}(\bm{\mu}_{{\bf Z}_{\bf B}},\bm{\Sigma}_{{\bf Z}_{\bf B}})$ with
   \begin{eqnarray}
    \label{post_Z_B}
    \begin{split}
    \bm{\Sigma}_{{\bf Z}_{\bf B}} &= [10^{-1}{\bf I}_{R_{\bf B}Q} + ({\Tilde{\bf X}}_{\bf B}^{\intercal} \otimes {\bf L}_{\bf B})^{\intercal} ({\bf I}_{(T-1)K} \otimes \bm{\Sigma}_{\bf B})^{-1}({\Tilde{\bf X}}_{\bf B}^{\intercal} \otimes {\bf L}_{\bf B})]^{-1}\\
    &= [10^{-1}{\bf I}_{R_{\bf B}Q} + ({\Tilde{\bf X}}_{\bf B}{\Tilde{\bf X}}_{\bf B}^{\intercal} \otimes {\bf L}_{\bf B}^{\intercal}\bm{\Sigma}_{\bf B}^{-1}{\bf L}_{\bf B})]^{-1},\\
  \bm{\mu}_{{\bf Z}_{\bf B}} &=  \bm{\Sigma}_{{\bf Z}_{\bf B}}  ({\Tilde{\bf X}}_{\bf B}^{\intercal} \otimes {\bf L}_{\bf B})^{\intercal} ({\bf I}_{(T-1)K} \otimes \bm{\Sigma}_{\bf B})^{-1} \text{vec}(\Tilde{{\bf Y}}_{\bf B})= \bm{\Sigma}_{{\bf Z}_{\bf B}}      \text{vec}({\bf L}_{\bf B}^{\intercal}\bm{\Sigma}_{\bf B}^{-1}\Tilde{{\bf Y}}_{\bf B}{\Tilde{\bf X}}_{\bf B}^{\intercal}),
    \end{split}
    \end{eqnarray}
while \eqref{eq:L_fac_B} leads to $(\text{vec}({\bf L}_{\bf B}) \mid -) \sim \mbox{N}_{R_{\bf B} Q}(\bm{\mu}_{{\bf L}_{\bf B}},\bm{\Sigma}_{{\bf L}_{\bf B}})$, where $\bm{\mu}_{{\bf L}_{\bf B}}$ and $\bm{\Sigma}_{{\bf L}_{\bf B}}$ are obtained as
      \begin{eqnarray}
    \label{post_L_B}
    \begin{split}
    \bm{\Sigma}_{{\bf L}_{\bf B}} &= [({\Tilde{\bf X}}_{\bf B}^{\intercal}{\bf Z}_{\bf B}^{\intercal} \otimes {\bf I}_Q)^{\intercal} ({\bf I}_{(T-1)K} \otimes \bm{\Sigma}_{\bf B})^{-1}({\Tilde{\bf X}}_{\bf B}^{\intercal}{\bf Z}_{\bf B}^{\intercal} \otimes {\bf I}_Q)]^{-1}\\
    &= ({\bf Z}_{\bf B}{\Tilde{\bf X}}_{\bf B}{\Tilde{\bf X}}_{\bf B}^{\intercal}{\bf Z}_{\bf B}^{\intercal} \otimes \bm{\Sigma}_{\bf B}^{-1})^{-1},\\
  \bm{\mu}_{{\bf L}_{\bf B}} &=   \bm{\Sigma}_{{\bf L}_{\bf B}} ({\Tilde{\bf X}}_{\bf B}^{\intercal}{\bf Z}_{\bf B}^{\intercal} \otimes {\bf I}_Q)^{\intercal} ({\bf I}_{(T-1)K} \otimes \bm{\Sigma}_{\bf B})^{-1} \text{vec}(\Tilde{{\bf Y}}_{\bf B})= \bm{\Sigma}_{{\bf L}_{\bf B}}  \text{vec}(\bm{\Sigma}_{\bf B}^{-1}\Tilde{{\bf Y}}_{\bf B}{\Tilde{\bf X}}_{\bf B}^{\intercal} {\bf Z}_{\bf B}^{\intercal}).
    \end{split}
    \end{eqnarray}
Finally, similarly to \eqref{post_Sigma_A}, the full conditional for $\bm{\Sigma}_{\bf B}$ given ${\bf B}= {\bf L}_{\bf B}{\bf Z}_{\bf B}$ is
\begin{eqnarray}
(\bm{\Sigma}_{\bf B} \mid -) \sim \mbox{inv-Wishart}(2 + (T-1)K, 2\cdot{\bf I}_Q +  ( \Tilde{{\bf Y}}_{\bf B} - {\bf B}{\Tilde{\bf X}}_{\bf B})( \Tilde{{\bf Y}}_{\bf B} - {\bf B}{\Tilde{\bf X}}_{\bf B})^{\intercal}).
    \label{post_Sigma_B}
\end{eqnarray}

Algorithm~\ref{alg:gibbs} summarizes the steps of the Gibbs sampler resulting from the above derivations. Notice that, due to the Kronecker structure of the matrix autoregressive model and the low-rank factorization of the coefficient matrices, some identifiability issues may arise. First, the matrix products $\mathbf{L}_{\bf A} \mathbf{Z}_{\bf A}$ and $\mathbf{L}_{\bf B} \mathbf{Z}_{\bf B}$ are only identifiable up to rotation, sign, scale and ordering of the respective rows and columns. Enforcing the constraints ${\bf L}_{\bf A}^{\intercal} {\bf L}_{\bf A}= {\bf I}_{R_{\bf A}}$ and ${\bf L}_{\bf B}^{\intercal} {\bf L}_{\bf B}= {\bf I}_{R_{\bf B}}$ discussed in Section~\ref{prior} ensures scale and rotational identification, and increases numerical stability of the algorithm. Such constraints are included in the Gibbs sampler via an intermediate processing step that first computes the QR decomposition  ${\bf L}_{\bf A}={\bf Q}_{L_{\bf A}}{\bf R}_{L_{\bf A}}$ and ${\bf L}_{\bf B}={\bf Q}_{L_{\bf B}}{\bf R}_{L_{\bf B}}$ of ${\bf L}_{\bf A}$ and ${\bf L}_{\bf B}$, respectively, and then redefines  ${\bf L}_{\bf A}={\bf Q}_{L_{\bf A}}$, ${\bf Z}_{\bf A}={\bf R}_{L_{\bf A}}{\bf Z}_{\bf A}$, and ${\bf L}_{\bf B}={\bf Q}_{L_{\bf B}}$, ${\bf Z}_{\bf B}={\bf R}_{L_{\bf B}}{\bf Z}_{\bf B}$. In this way, ${\bf A}={\bf L}_{\bf A}{\bf Z}_{\bf A}$ and ${\bf B}={\bf L}_{\bf B}{\bf Z}_{\bf B}$ are unchanged, while meeting the identifiability constraints ${\bf L}_{\bf A}^{\intercal} {\bf L}_{\bf A}= {\bf I}_{R_{\bf A}}$ and ${\bf L}_{\bf B}^{\intercal} {\bf L}_{\bf B}= {\bf I}_{R_{\bf B}}$. Column and sign switching was not an issue in our case, but  could be resolved via order constraints in the \textsc{mcmc} or ex post identification strategies. 

Finally, recalling results in  \citet{hoff2015multilinear}, ${\bf A}$, ${\bf B}$ and ${\bf \Sigma}_{\bf A}$, ${\bf \Sigma}_{\bf B}$ are only identified up to a scaling, i.e.\ ${\bf B} \otimes {\bf A}=({\bf B}/c) \otimes (c{\bf A})$ and  ${\bf \Sigma}_{\bf B} \otimes {\bf \Sigma}_{\bf A}=({\bf \Sigma}_{\bf B}/c) \otimes (c{\bf \Sigma}_{\bf A})$. To address this issue, we follow the recommendation in Appendix B of  \citet{hoff2015multilinear} by rescaling ${\bf A}$, ${\bf B}$ and  ${\bf \Sigma}_{\bf A}$, ${\bf \Sigma}_{\bf B}$ so as to maintain a constant relative magnitude among $||{\bf A}||^2$ and $||{\bf B}||^2$, and between  $||{\bf \Sigma}_{\bf A}||^2$ and $||{\bf \Sigma}_{\bf B}||^2$, respectively.

\section{\large 5. Modeling the criminal activity tensor in the city of Chicago} \label{sec:application}
We proceed by illustrating the predictive performance and the inference potential of the model proposed in Sections~\ref{sec:model} and \ref{sec_3}, with a focus on criminal activity data in Chicago. These data are described  in Section~\ref{sec:data}. The results of the analysis  can be found in  Section~\ref{sec:results}.

\begin{figure}[b]
    \centering
    \includegraphics[width=0.88\textwidth]{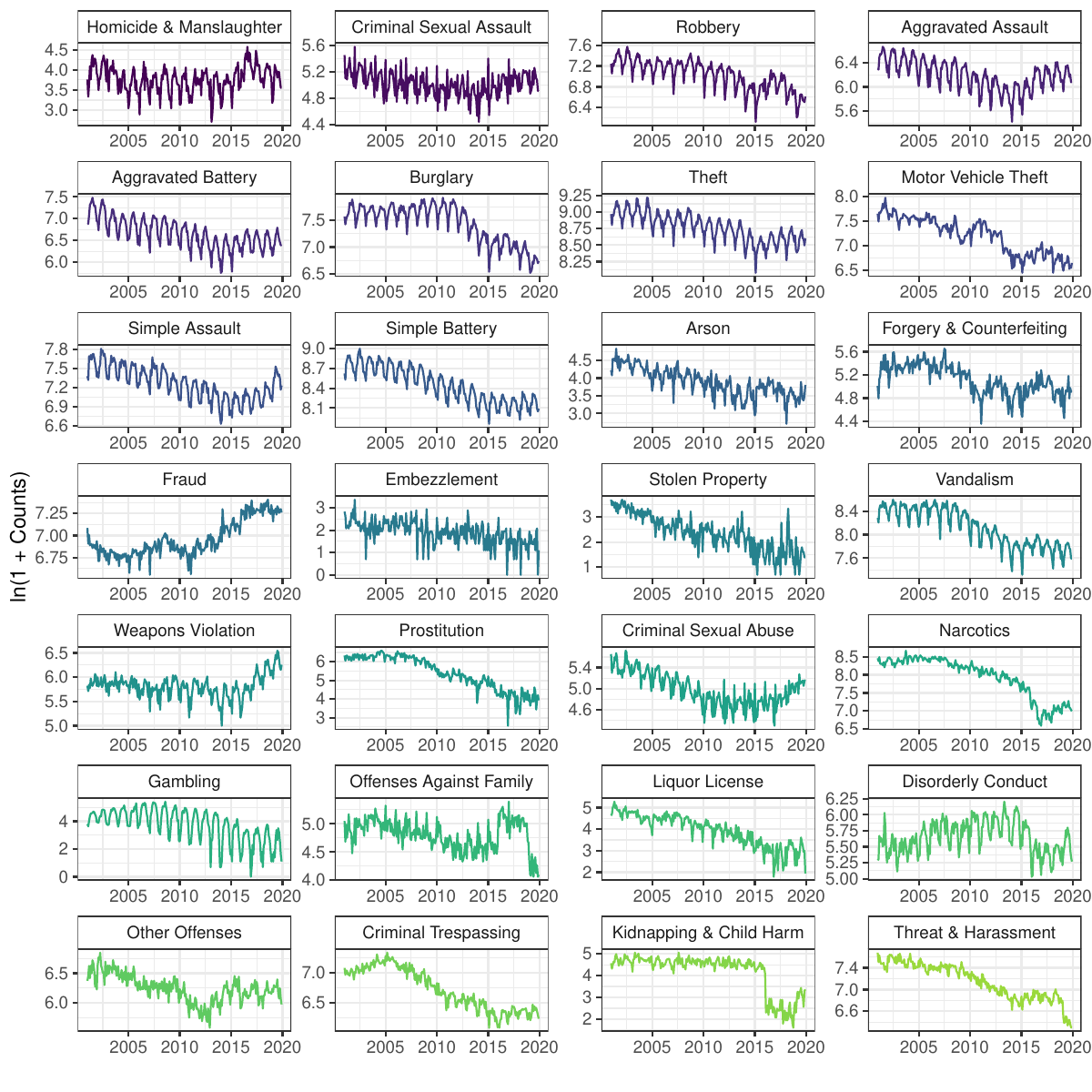}
    \caption{\footnotesize For each crime category, log-transformed counts of police reports by month, aggregated across districts. The time window spans from January 2001 to December 2019. }
    \label{fig:log_counts}
\end{figure}

\subsection{\large 5.1. Data}  \label{sec:data}
\begin{figure}[t]
     \centering
     \begin{subfigure}[b]{0.29\textwidth}
         \centering
 \includegraphics[width=\linewidth]{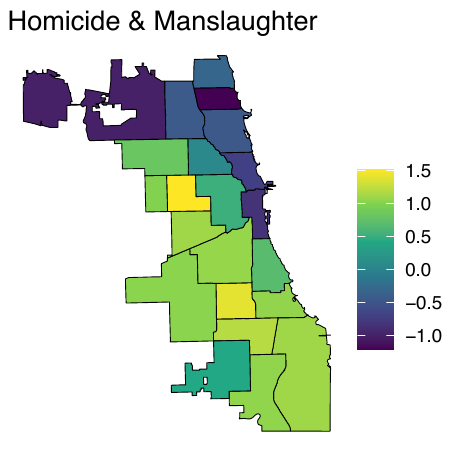}
     \end{subfigure}
     \hfill
     \begin{subfigure}[b]{0.29\textwidth}
         \centering
 \includegraphics[width=\linewidth]{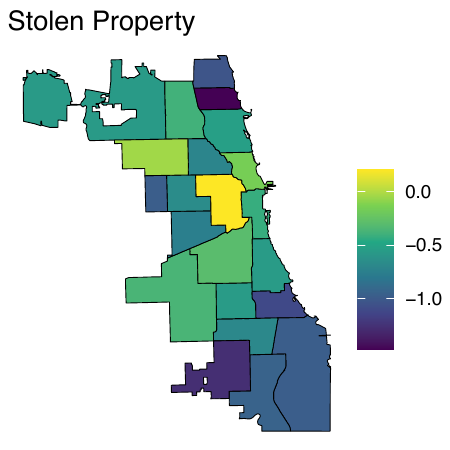}
     \end{subfigure}
          \hfill
     \begin{subfigure}[b]{0.29\textwidth}
         \centering
 \includegraphics[width=\linewidth]{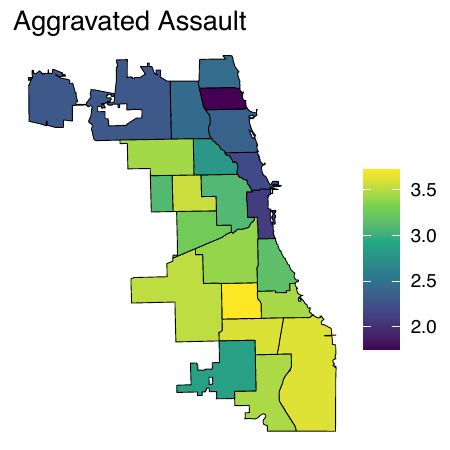}
          \end{subfigure}
        \caption{\footnotesize Heatmaps of reports' log-transformed counts for three selected crime categories in each district of Chicago, averaged over time. }
        \label{fig:barrio_examples}
\end{figure}

The three-way criminal activity tensor we rely on is available at \url{https://www.chicagopolice.org/data-statistics/} and comprises time-stamped reports from the Chicago Police Department on $K=28$ crime categories within the $Q=22$ districts of the city. In our application, these data are aggregated on a monthly timescale from January 2001 until December 2019 in order to avoid structural breaks due to the \textsc{covid}-19 pandemic in the early 2020. Figure~\ref{fig:log_counts} shows the dynamic log-counts of these reports for each crime category, aggregated over districts, while Figure~\ref{fig:barrio_examples} displays selected heatmaps of the log-transformed number of crime reports across the $Q=22$ districts for specific categories, averaged over the time window analyzed. Consistent with the discussion in Section~\ref{sec:lit}, both figures clarify that criminal activities display heterogeneous structures across time, crime categories and districts, which further showcase  complex interdependencies motivating the implementation of the model developed in  Sections~\ref{sec:model}--\ref{sec_3}. As discussed in Section~\ref{sec:model}, this model is applied to the detrended and seasonally adjusted version of the time series in Figure~\ref{fig:log_counts} (refer to Figure~B1 in the Supplementary Material).

Note that the $K=28$ crime categories studied  mainly follow the classification system employed by the Federal Bureau of Investigation (\textsc{fbi}), which also aligns with the National Incident-Based Reporting System (\textsc{nibrs}). However, we observed that a \textsc{nibrs} category \texttt{Miscellaneous Other} often played an intriguing role in our results. To enhance the interpretability, we therefore decided to redistribute the crimes within this category based on the corresponding Illinois Uniform Crime Reporting (\textsc{iucr}) code (see the Supplementary Material). For example, we reallocated reports corresponding to \textsc{iucr} code {\em 2092 -- solicit narcotics on public way} to the \textsc{fbi} category \texttt{Drug Abuse (18)}, which therefore represents a broader, narcotics-related category. Furthermore, certain other \textsc{iucr}  codes previously part of \textsc{fbi} category \texttt{Miscellaneous Other} have been bundled into new, separate, crime categories. An example for this are several codes related to threat and harassment, for instance {\em 2820 - telephone threat} or {\em 2826 - harassment by electronic means}. We reclassified these codes into an explicit \texttt{Threat \& Harassment} category. This refinement of the crime classification allows us to capture the nuances of different crimes more accurately.

Before discussing the results of our analysis, it is also worth mentioning that any study based on police reports provides only a partial understanding of the underlying criminal activity patterns due to several factors. First, the police reports inherently encode information on criminal activity filtered through the lens of law enforcement. As such, it is important to acknowledge that the reported incidents may not perfectly align with the actual criminal activities across the city. Second, the spatial dimension of these reports introduces complexities, as different neighborhoods may have varying levels of trust in the police, leading to differences in reporting patterns. Finally, the type of crimes being reported and the temporal aspect can also impact the results, further highlighting the need for caution when interpreting the findings. Nonetheless, despite these limitations, analyzing police reports remains the standard approach in state-of-the-art literature, and is commonly used for forecasting and utilized by the law enforcement itself. It provides valuable insights into crime patterns, police activity, and the structures that emerge from the interaction among the two. That is, the results may be influenced by both the system of policing and the actual criminal activity, making it challenging to disentangle the two. Nonetheless, the interpretable model we propose can uncover hidden patterns in the data produced by law enforcement, thus providing police with more refined methods to  leverage information from these reports.

\subsection{\large 5.2. Results}
\label{sec:results}
The presentation of the results of the analysis under the proposed model is divided in four parts. In Section~\ref{predictive}, we first compare the forecasting accuracy of model  \eqref{eq:main_eq}--\eqref{eq:fact} with those achieved by a full-rank bilinear autoregressive formulation and a black-box predictive strategy relying on random forests. Besides clarifying that the proposed parsimonious representation is competitive with strategies specifically designed for forecasting, this assessment provides reassurance on the fact that inference on the dependence structures encoded in $\bm{\Sigma}_{\bf A}$, $\bm{\Sigma}_{\bf B}$, ${\bf A}$ and ${\bf B}$ relies on a realistic characterization of the observed criminal activity tensor. The results of inference on the posterior point estimates for these quantities are discussed in  Sections~\ref{sec:intra} and \ref{sec:inter}, respectively.  Finally, in Section~\ref{sec:find} we interpret the findings arising from such inferences in the light of the  qualitative and quantitative evidence on criminal activities presented in Section~\ref{sec:lit}.

All the results presented leverage Monte Carlo estimates based on $3{,}000$ posterior samples for the quantities of interest produced by Algorithm~\ref{alg:gibbs}, following a burn-in period of $1{,}000$. The total runtime of this routine is 20 minutes on a standard laptop, much lower than the monthly time scale at which data are analyzed, and hence, the estimates would require updating. Convergence and mixing of  Algorithm~\ref{alg:gibbs} posed no major challenges. To select  $R_{\bf A}$ and $R_{\bf B}$, we conducted an exhaustive enumeration of all  possible values for $(R_{\bf A},R_{\bf B})$, evaluating these combinations in terms of out-of-sample predictive accuracy, similarly to the assessments conducted in Section~\ref{predictive}. The results of this analysis suggest  $R_{\bf A} = 11$ and $R_{\bf B} = 5$. This provides a parsimonious specification for model  \eqref{eq:main_eq}--\eqref{eq:fact} that bounds the loss in predictive accuracy  (measured by out-of-sample root mean squared error and mean absolute error) to approximately 1\% of the full-rank formulation.

\subsubsection{5.2.1. Predictive performance and benchmarking} \label{predictive}
Before discussing the results of inference under model in \eqref{eq:main_eq}--\eqref{eq:fact}, let us first assess its forecasting performance and compare it against alternative predictive strategies. In presenting the results of this analysis, it is important to emphasize that while a significant portion of the literature on crime data analysis is primarily designed for purely predictive purposes, the model developed in Sections~\ref{sec:model}--\ref{sec_3} aims instead at providing interpretable inference on the dependence structures within the observed criminal activity tensor. As such, the overarching goal of our predictive assessment is not to demonstrate that the proposed model systematically outperforms all potential competitors, including black-box ones, in out-of-sample forecasting. Rather, we aim to provide empirical evidence on the fact that carefully designed model-based representations, such as the proposed one, can substantially enhance interpretability and inference potentials, while preserving a predictive accuracy that is competitive with those achieved by  highly parameterized  formulations and black-box solutions. This assessment also  serves as a sanity check to quantify whether the proposed model yields a realistic representation of the generative mechanism underlying the data analyzed, thereby reassuring on the reliability of the inference results  in  Sections~\ref{sec:intra}--\ref{sec:inter}.

\begin{table}[b]
\centering
\caption{\footnotesize Forecasting performance of the proposed model (low-rank bilinear autoregression) and relevant competitors (full-rank bilinear  autoregression, random forests) under different metrics: \textsc{rmse} (root mean squared error), \textsc{mae} (mean absolute error) and \textsc{corr} (correlation between predictions and actual observed values). Results are averaged across 24 replicated one-step-ahead forecasting exercises. The low-rank bilinear autoregressive model employs $R_{\bf A} = 11$, $R_{\bf B} = 5$.}
\begin{tabular}{lccc}
 & \textsc{rmse} & \textsc{mae} & \textsc{corr} \\
  \midrule
Low-rank bilinear autoregression & 0.374 & 0.269 & 0.975   \\
\midrule
Full-rank bilinear  autoregression \qquad  \qquad  & 0.370 & 0.267 & 0.975   \\
Random forests  & 0.365 & 0.270 & 0.976   \\
    \midrule
\end{tabular}
\label{tab:accuracy}
\end{table}

Consistent with the above goals, we study the predictive performance of the proposed model (low-rank bilinear autoregression with $R_{\bf A} = 11$, $R_{\bf B} = 5$) and two relevant competing methods. The first one (full-rank bilinear autoregression) still relies on the bilinear representation as in \eqref{eq:main_eq}, but does not include the low-rank structure in  \eqref{eq:fact}. The second (random forests \citep{breiman2001random}) provides instead a relevant example of a routinely implemented black-box strategy specifically designed for predictive purposes. In evaluating these competing strategies, we divide the data into a training set and a test set. The training set covers data for 204 consecutive months from $t=1$ until $t=t_{\mbox{\tiny train}}$. Leveraging these training data, we fit all the three methods under analysis, including the proposed model, and then produce one-step-ahead forecasts for the log-transformed criminal activity counts at $t_{\mbox{\tiny train}}+1$ for all the 616 combinations of districts and crime categories. Recalling Section~\ref{sec:model}, under the proposed model and its full-rank version the predictions for the original log-transformed counts  can be readily obtained by adding the deterministic component $\hat{\mu}_{kq(t_{\mbox{\tiny train}}+1)}$ to the forecasts for $y_{kq(t_{\mbox{\tiny train}}+1)}$ produced by these two strategies. The performance of random forests is instead assessed by training such a method directly on the observed log-counts $\log(1+c_{kqt})$, while leveraging as predictors the time index $t$, its square, the seasonal dummies ${\bf d}_t$ (see Section~\ref{sec:model}) and the lagged  log-counts. This yields a flexible formulation, which can learn even more complex systematic components than those we rely for detrending and seasonal adjustments. Note that, consistent with this competitor and with routinely-adopted black-box approaches, we focus on point forecasts also under our model and leave quantification of predictive uncertainty for future research. This  perspective would require a careful treatment  to properly propagate the uncertainty from the detrending and seasonal adjustment considered in Section~\ref{sec:model}.

\begin{figure}[t]
    \centering
    \includegraphics[trim={0cm 0cm 0 0cm},clip,width = 0.85\textwidth]{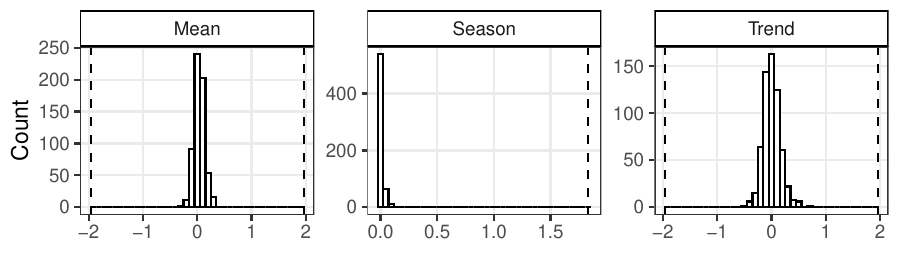}
 \caption{\footnotesize Distribution of residual diagnostics across all crime-district series: (left) $z$-scores for mean residuals, (middle) $F$-statistics testing for 12-month seasonality, and (right) $z$-scores for linear trends. Dashed vertical lines indicate the 5\% critical values ($\pm 1.96$ for mean and trend, and the 5\% $F$-threshold for seasonality). The distributions are tightly centered within the thresholds, indicating satisfactory residuals behavior.}
    \label{fig:resid}
\end{figure}

The forecasts produced by the above methods are compared to the observed values of the  log-transformed counts  at $t_{\mbox{\tiny train}}+1$ using three accuracy measures, namely the root mean squared error (\textsc{rmse}), the mean absolute deviation (\textsc{mae}) and the correlation (\textsc{corr}) between the forecasted and observed values. To provide a comprehensive assessment, these measures are averaged over 24 replications of such a predictive analysis, each obtained by shifting the training window of 204 observations forward by one month, and forecasting  in the subsequent month. As a result, predictive performance is quantified on a total of $616\cdot 24=14{,}784$ point forecasts. Table~\ref{tab:accuracy} summarizes the outcome of this study. Interestingly, although one would expect random forests to substantially improve the forecasting accuracy of the low-rank and full-rank bilinear autoregressive models, this is not the case in such a context. In fact, both model-based solutions achieve a predictive accuracy that is comparable to random forests, with the low-rank formulation proposed in~\eqref{eq:main_eq}--\eqref{eq:fact} providing essentially the same performance as its full-rank counterpart. 

The above results clarify that interpretable inference can be accomplished at a negligible cost in predictive accuracy, and provides reassurance that the findings in  Sections~\ref{sec:intra}--\ref{sec:inter} are  not based on an unrealistic  characterization of the data-generating mechanism. This point is further underscored by  Figure~\ref{fig:resid}, which shows that the residuals resulting from our model do not display, for all crime-district series, evidence of non-zero mean, trends or seasonal effects. Such a finding, combined with the results in Table~\ref{tab:accuracy}, also supports the adequacy  of the  series-by-series detrending and seasonal adjustment considered in Section~\ref{sec:model}.

\subsubsection{5.2.2. Inference on intra-temporal dependence structures} \label{sec:intra}
\begin{figure}[b]
    \centering
    \includegraphics[trim={0cm 4cm 0 5cm},clip,width = 0.97\textwidth]{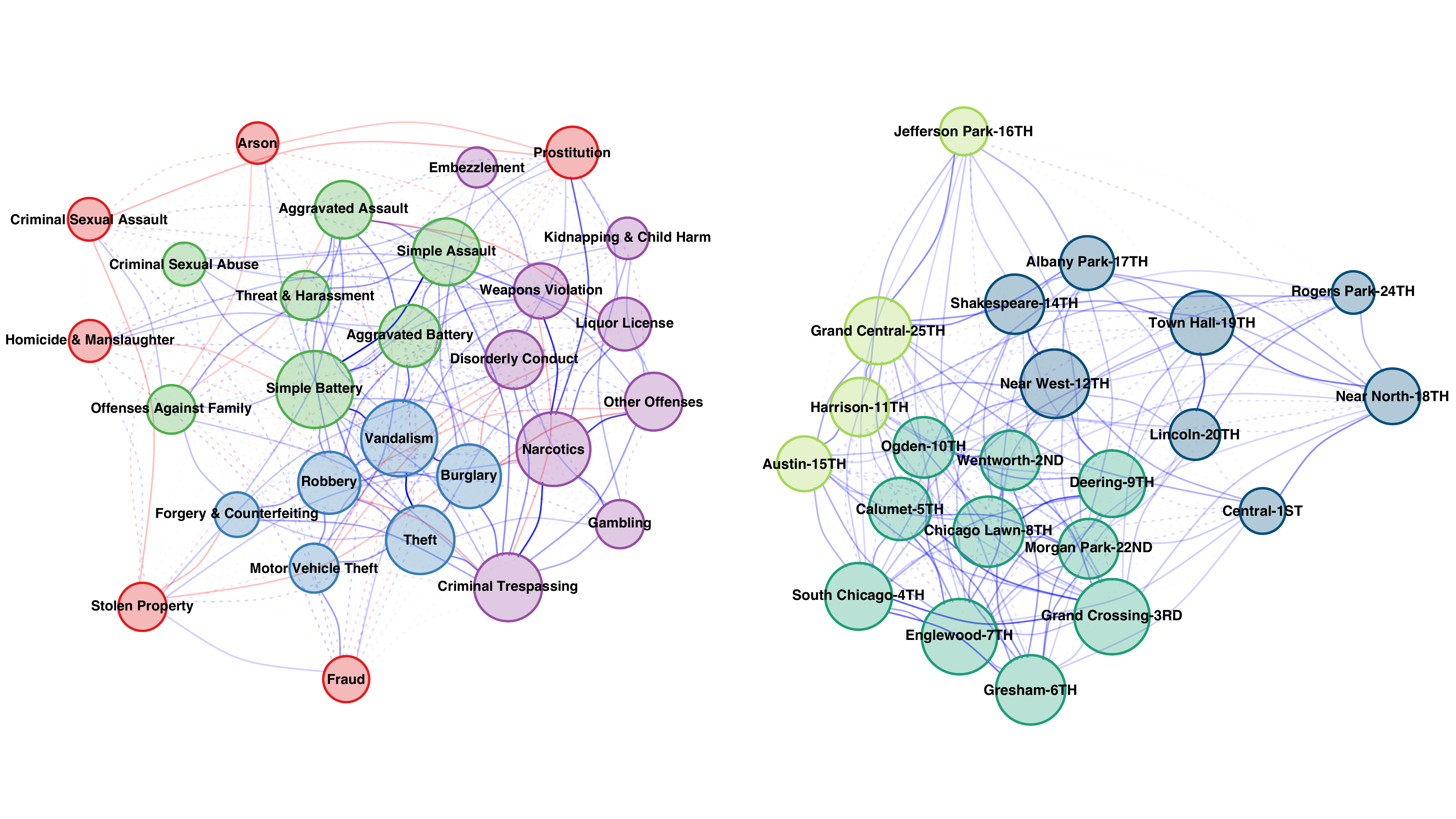}
 \caption{\footnotesize Estimated  intra-temporal partial correlation networks between crime categories (left) and among districts (right). The colors in each network indicate clusters of nodes with high within-group absolute partial correlations and relatively low across-group dependencies (clusters are estimated via community detection algorithms \citep{blondel2008fast}). Node positions are obtained via force-direct placement methods \citep{fruchterman1991graph} positioning closer in the space nodes with higher absolute partial correlations, whereas node size is proportional to the sum of the absolute partial correlations involving that node. Red and blue lines denote negative and positive partial correlations, respectively, while the degree of shading of each line goes from maximum transparency to no transparency as the associated absolute partial correlation moves from zero to the maximum value within each network  (the estimated partial correlations range from   $-0.05$ to $0.20$ for the crime category network and from   $-0.02$ to $0.09$ for the district network). Finally, solid (dashed) lines refer to partial correlations whose induced posterior has $90\%$ credible intervals not including (including) the value $0$.}
    \label{fig:networks}
\end{figure}

Besides achieving satisfactory predictive performance, a distinctive feature of the model developed in Section~\ref{sec:model}--\ref{sec_3} lies in its ability to disentangle complex dependencies within the criminal activity tensor, while allowing interpretable inference on such quantities. To showcase such potentials, let us first focus on the intra-temporal dependence structures among crime categories and between districts encoded in the posteriors  of $\bm{\Sigma}_{\bf A}$ and $\bm{\Sigma}_{\bf B}$. Figure~\ref{fig:networks} provides a graphical representation of these estimated dependencies with a focus on the posterior mean of the partial correlation networks defined as $- \mbox{diag}(\bm{\Sigma}_{\bf A}^{-1})^{-1/2}\bm{\Sigma}_{\bf A}^{-1}\mbox{diag}(\bm{\Sigma}_{\bf A}^{-1})^{-1/2}$ and $- \mbox{diag}(\bm{\Sigma}_{\bf B}^{-1})^{-1/2}\bm{\Sigma}_{\bf B}^{-1}\mbox{diag}(\bm{\Sigma}_{\bf B}^{-1})^{-1/2}$ \citep[e.g.][]{lauritzen1996graphical}. Such posterior means can be obtained by evaluating these formulas at the samples of  $\bm{\Sigma}_{\bf A}$ and $\bm{\Sigma}_{\bf B}$ produced by Algorithm~\ref{alg:gibbs} and then averaging across samples. Crucially, these measures provide an informative reconstruction of  graphical dependencies between crime categories and across the different  districts, which  quantify the intra-temporal pairwise correlations among any two generic nodes in the network conditionally on the effect of all the others, and  net of the inter-temporal dependencies captured by the conditional mean in~\eqref{eq:main_eq}--\eqref{eq:fact}. 

 Figure~\ref{fig:networks} displays the aforementioned networks with nodes corresponding to crime categories and districts, respectively, and weighted edges denoting  the posterior mean of the partial correlation among any pair of nodes. To facilitate interpretation and uncover relevant structures, the positions of the nodes within  Figure~\ref{fig:networks} are obtained via force-direct placement methods \citep{fruchterman1991graph} so that nodes with higher absolute partial correlations are closer in the space. This graphical setting is also combined with the output of a community detection algorithm \citep{blondel2008fast}, which unveils clusters of nodes having high within-group absolute partial correlations and relatively low across-group dependencies. As shown in the network among the crime categories, such a perspective allows us to uncover yet unexplored modular intra-temporal dependencies between types of crimes. More specifically, the first group (green nodes) comprises, among others, crimes such as simple/aggravated battery and assault, which are physical crimes directly targeting the body. The second (purple nodes) includes, instead, crimes related to narcotics, gambling, weapons violations, and liquor license-related offenses. This clustering suggests a potential association with organized crime or coordinated gang activities in the city. The third group (blue nodes) consists mainly of property crimes such as burglary, robbery, theft and vandalism. These  crimes are often interconnected for various reasons. For example, evidence suggests that the presence of vandalism, like graffiti, can significantly increase the propensity of people to engage in property crimes, such as stealing \citep[][]{keizer2008spreading}.  Finally, the last group (red nodes) captures crimes that exhibit sparser dependencies, even within the community itself. This cluster encompasses crimes such as arson, homicide or fraud, which are offenses that tend to occur independently of other crimes. A partial exception in such a group is provided by prostitution, which appears in  Figure~\ref{fig:networks} relatively close to the purple nodes, thereby suggesting that this category might have also direct connections with organized crime or coordinated gang activities. Besides the  composition of each group, it is also worth noticing how simple battery, vandalism and narcotics play a central role within the partial correlation network, while being allocated to three different crime communities. This finding suggests that policies simultaneously targeting these three crime categories across the whole city could trigger positive spillover effects on other crime types. See Section~\ref{sec:find} for further discussion on these results in the light of the qualitative and quantitative criminology theories presented in Section~\ref{sec:lit}.

\begin{figure}[b]
    \centering
    \includegraphics[trim={1.4cm 0 0cm 0cm},clip,width = 0.8\textwidth]{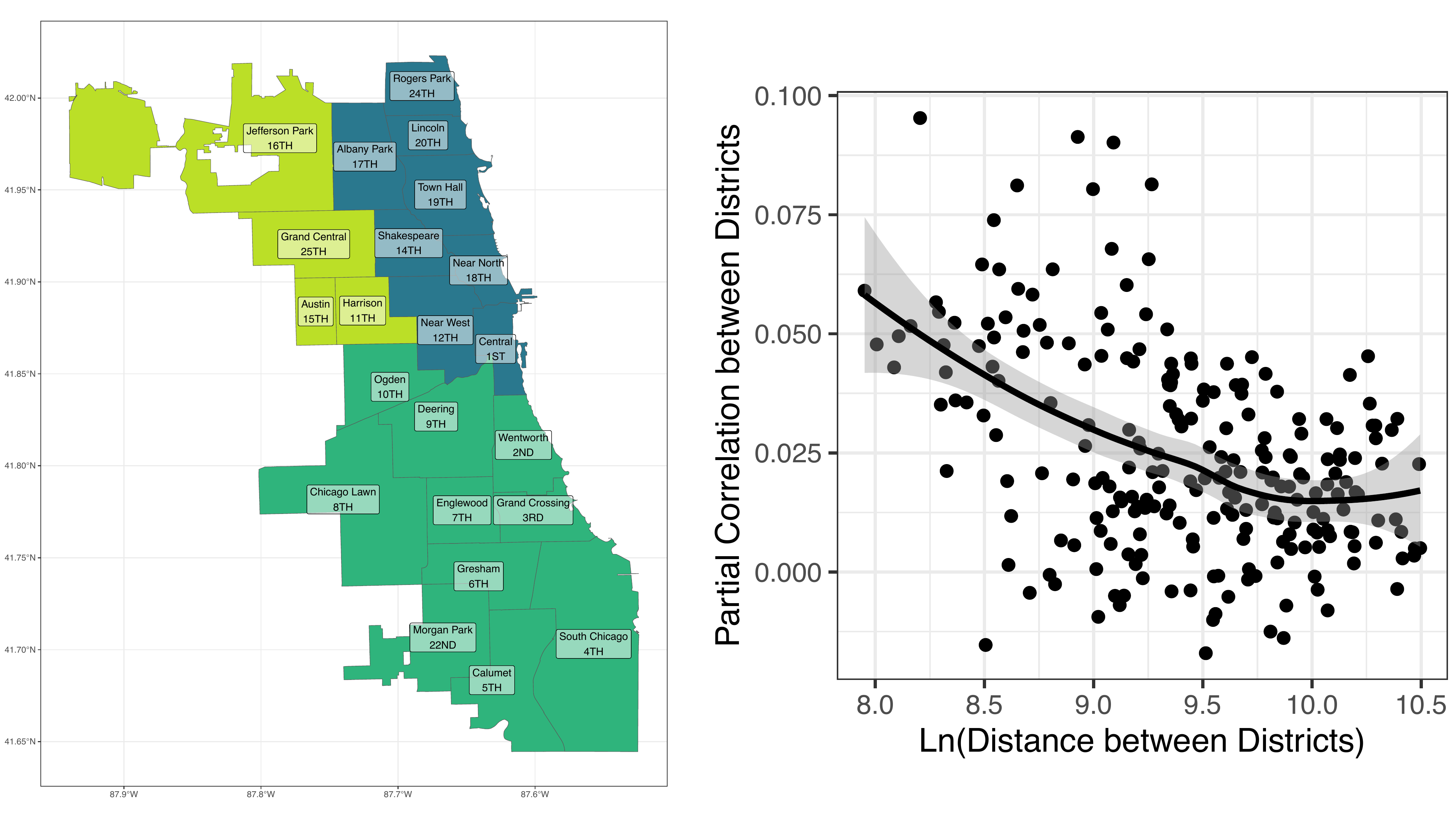}
    \caption{\footnotesize Left: Districts of Chicago colored according to the communities estimated from the partial correlation network in the right panel of  Figure~\ref{fig:networks}. Right: Plot of log-transformed distances between pairs of district centroids versus the posterior mean of the partial correlation among such district pairs (\textsc{loess} estimate and uncertainty bands for the relation between these two quantities are also provided).}
    \label{fig:district_communities}
\end{figure}

Similarly interesting results can be obtained by examining the network structure of the partial correlations among districts  in the right panel of Figure~\ref{fig:networks}.  The community detection algorithm \citep{blondel2008fast} applied to this network reveals  three distinct groups, which align with the north-east, north-west and south division of  Chicago. This geographical separation becomes even more evident when visualizing the estimated group structure on a map of the city, as shown in the left panel of Figure~\ref{fig:district_communities}. The connection of the partial correlations to geographical distance is further underlined in the right panel of  Figure~\ref{fig:district_communities}, which highlights how the overall negative relationship between the pairwise distances among districts and the corresponding partial correlations is combined with heterogeneous patterns that partially depart from pure proximity structures. As discussed in Section~\ref{sec:find}, these findings contribute to the ongoing studies on criminal activities in urban environments and provide useful resources to quantify potential spillover effects across districts. Remarkably, the proposed formulation identifies spatial dependence structures without including  explicit geographical information within the model.  In fact, as anticipated in Section~\ref{prior}, avoiding these constraints allows us to learn in the right panel of Figure~\ref{fig:networks} stronger dependencies within southern and north-western districts relative to those inferred among north-eastern ones. This result partially departs from  pure proximity structures that would have been imposed by, e.g.\ \textsc{car} constructions or related spatio-temporal formulations \citep[e.g.][]{banerjee2003hierarchical,botella2015unifying,vicente2023multivariate},  and provides
 quantitative evidence that supports the reported presence of organized crime and structured gangs activities in the southern  and north-western area of Chicago  \citep[e.g.][]{aspholm2019views,schnell2017influence}. This finding is  confirmed by the size of the districts in Figure~\ref{fig:networks}, which indicates that southern and north-western areas play a key role in the diffusion of crime across the city. Notice that while the partial correlations presented in  Figure~\ref{fig:networks} (and Figure~\ref{fig:district_communities}) are modest in magnitude (an unsurprising consequence of the pre-processing employed and the relatively flexible conditional mean structure in~\eqref{eq:main_eq}), the posteriors for several of these partial correlations concentrate on non-zero values. This suggests the presence of informative residual dependence that supports the adoption of a structured representation rather than an independent error model.

\subsubsection{5.2.3. Inference on inter-temporal dependence structures} \label{sec:inter}
\begin{figure}[b]
    \centering
    \includegraphics[trim={0cm 6cm 0 6.5cm},clip,width =\textwidth]{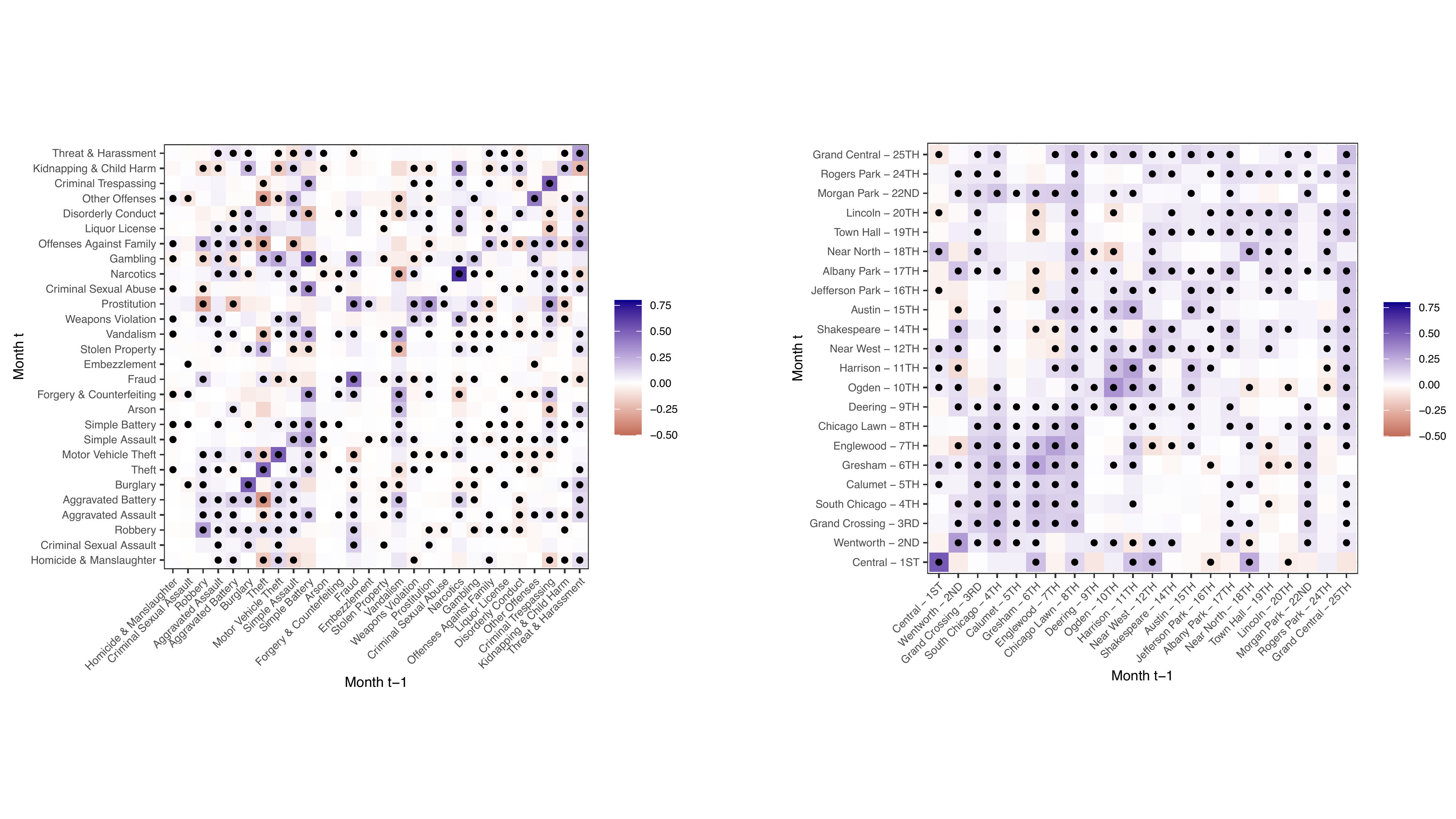}
    \caption{\footnotesize Heatmaps of the posterior mean for the matrices ${\bf A}$ (left) and ${\bf B}$ (right) characterizing the inter-temporal dependencies under model \eqref{eq:main_eq}--\eqref{eq:fact}. The black dots highlight parameters whose posterior distribution has $90\%$ credible intervals not including the value $0$.}
    \label{fig:var_crimes}
\end{figure}

Figure~\ref{fig:var_crimes} displays the heatmaps of the posterior means for the elements in the matrices ${\bf A}$ and ${\bf B}$. Recalling Section~\ref{sec:formul}, these quantities  provide key insights into the shared inter-temporal dependence among the $K$ crime categories within districts, and between the $Q$ districts for each crime category, respectively. As such,  although these estimates may not directly correspond to causally-interpretable effects,  Figure~\ref{fig:var_crimes}   can still unveil relevant association patterns, which complement those presented within Section~\ref{sec:intra}. Recall that intra- and inter-temporal dependencies have a substantially different meaning. In fact, crime categories (or districts) displaying instantaneous intra-temporal effects, may not necessarily influence each other across different time lags.

According to  Figure~\ref{fig:var_crimes}, the dominant feature of both matrices is the main diagonal, indicating a moderately high inter-temporal persistence within reports by crime categories and districts, respectively. In the posterior mean of the matrix  ${\bf A}$ (see left panel of Figure~\ref{fig:var_crimes}) certain crimes, such as arson and embezzlement occur less frequently and in relative isolation, leading to exceptions in this overall pattern. Additionally, the occurrence of specific crime categories, such as property- and aggression-related crimes, appears to have a more remarkable significant association with the potential outbreak of specific types of crimes in the subsequent month. Such a finding provides important guidelines in the design of intervention policies guided by potential spillover effects.

Relevant findings are provided also by the posterior mean of the matrix  ${\bf B}$ in the right panel of 
Figure~\ref{fig:var_crimes}. Besides the direct inter-temporal persistence highlighted in the main diagonal, the estimated ${\bf B}$ clarifies that this persistence acts also at a group level, especially for the clusters of southern districts whose crime activities appear to be associated with those at subsequent times in other districts from the same group. In this group it is also possible to identify districts (e.g.\ Chicago Lawn) with significant inter-temporal effects on the entire city. Similar impact is found also for Grand Central, situated in the north-western area. These results are in line with the spatial intra-temporal dependencies discussed in Section~\ref{sec:intra} and further support the idea that these patterns might arise from the presence of organized crime and coordinated gangs activities in the southern and north-western area of the city   \citep[][]{aspholm2019views,schnell2017influence}.

\subsubsection{5.2.4. Discussion of the results} \label{sec:find}
The results in Sections~\ref{sec:intra}--\ref{sec:inter} provide novel empirical evidence in the context of criminological analyses on the mechanisms driving concentration of specific crime types in time and space  \citep[e.g.][]{brantingham2016geometry, brantingham2017notes, cohen1979social}. Social, economic, and even structural characteristics of urban environments influence criminal behavior \citep{brantingham1995criminality}, leading to specific inter- and intra-temporal dependencies within both crime categories and spatial districts that also exhibit peculiar modular patterns (see Figures~\ref{fig:networks} and \ref{fig:var_crimes}). Therefore, our results underscore the importance of maintaining a systemic perspective on crime phenomena, accounting for all the three dimensions of the criminal activity tensor, while inferring interpretable dependencies within and across such dimensions. For example, the spatial effects we unveil in Figures~\ref{fig:networks} and \ref{fig:var_crimes} highlight an heterogeneous distribution of crime generators and crime attractors within the urban environment \citep{brantingham1995criminality,weisburd2015law}, which, in turn, do not facilitate the same crime categories  across the city. An interesting case is provided by the Chicago Loop in the 1st district. This area is a financial, cultural, and commercial hub, thereby increasing the number of potential targets, especially in relation to property crime. Consistent with this, the right panels of  Figures~\ref{fig:networks} and \ref{fig:var_crimes} show that such a district is a crime generator area showcasing non-negligible intra- and inter-temporal associations with some southern and north-western districts. Recalling, \citet{aspholm2019views} and \citet{schnell2017influence}, these southern and north-western districts tend to act, instead, as crime attractors due to reduced interventions and structural organizational issues that create opportunities for organized criminal activities, and hinder residents from exerting informal social control \citep[][]{sampson1989community, shaw1942juvenile}. These criminogenic disparities across the different areas of Chicago seem to emerge in our findings, which highlight modular  intra- and inter-temporal associations among districts, with southern and north-western areas displaying more dense patterns.

The results in Figure~\ref{fig:district_communities} further underscore the importance of the spatial proximity and spillover effects. According to our findings, there is an overall negative relation between the distance among districts and the corresponding partial correlation. A possible explanation involves mechanisms that combine residential location and social interactions \citep{zenou2003spatial}, highlighting the role of behavior in disadvantaged areas \citep{glaeser1996crime}. Evidence suggests that offenders tend to exploit criminal opportunities close to their residence \citep[e.g.][]{bernasco2010sentimental, bernasco2009offenders,  hipp2016general, menting2016family, townsley2016target}. Consequently, the concentration of offenders in a neighborhood can trigger contagion effects \citep{becker2009social}.  

Moreover, increased criminal concentration intensifies the competition among offenders, reducing the expected gain of criminal activity \citep{becker1968crime} and encouraging repeat offenses \citep{zenou2003spatial}. This is also consistent with research on repeat victimization and temporal clustering, which indicates that the occurrence of an initial crime increases the probability of subsequent offenses in the same area. Offenders tend to exploit previously acquired knowledge about susceptible targets and absence of  interventions \citep[][]{johnson2004burglary, townsley2003infectious}. This crime persistence is also evident in Figure~\ref{fig:var_crimes}, although there are differences across categories. Specifically, while the main diagonal of the left matrix in Figure~\ref{fig:var_crimes}  shows temporal persistence for  property- and aggression-related crimes, as well as for organized crime offenses like narcotics, prostitution, and gambling, this pattern is less pronounced for violent offenses such as sexual-related crimes, homicides and manslaughter, and for property crimes like arson and embezzlement. These heterogeneities  provide novel empirical evidence to the literature on such phenomena \citep{chainey2021comparison, haberman2012predictive, johnson2004burglary, johnson2009offender}.

\section{\large 6. Conclusion}
\label{sec:concl}
This article proposes a low-rank bilinear autoregressive model to uncover interpretable dependence structure within and across the three dimensions (time, space, crime categories) of observed criminal activity tensors, with a specific focus on report data from the Chicago city. Besides disentangling and facilitating inference on complex interdependence patterns over several dimensions, the proposed formulation crucially achieves also a predictive accuracy that is competitive with black-box strategies, thereby providing a realistic, yet interpretable, model-based representation of crime phenomena over space, time and different categories. The results in the application confirm these advantages, and further stress the importance of developing joint models that maintain a systemic perspective on crime phenomena by formally accounting for dependencies across and within all the three dimensions of a generic criminal activity tensor. Such a perspective not only contributes to enlarging available knowledge on the mechanisms and determinants of crime, but also facilitates the design and evaluation of improved policies and intervention strategies.

The above advancements motivate several directions for future research. First, extending the model to incorporate nonlinearities could provide further insights into the complex dynamics of criminal activities that might also yield further gains in predictive accuracy. Relatedly, another interesting direction would be to explicitly incorporate additional structural information, such as knowledge on activity patterns of organized criminal groups or spatial structure, into the modeling framework. The latter could be achieved by introducing external covariates or by incorporating  spatio-temporal dependencies into the model parameters, leveraging, e.g.\ the results in  \citet[][]{hsu2021matrix} and \citet{martinez2016towards}, among others. Although the proposed model learns both distance-related effects and  departures from pure proximity patterns, additional empirical comparisons against these spatio-temporal alternatives could further illustrate the potential of our formulation. Along these lines, and consistent with the discussions in Sections~\ref{sec:intra}--\ref{sec:inter}, it would be interesting to formally compare competing error models (e.g.\ residuals' independence,  \textsc{car} priors and more flexible mechanisms as implied by our approach) via principled model comparison strategies such as proper scoring rules. Finally, given the typical dominance of the main diagonal in the estimated matrices ${\bf A}$ and ${\bf B}$, future research could consider  alternatives to the factorization  in \eqref{eq:fact}, such as, e.g. ${\bf A} = {\bf D}_{\bf A} + {\bf L}_{\bf A}{ \bf Z}_{\bf A}$ and ${\bf B} = {\bf D}_{\bf B} + {\bf L}_{\bf B}{ \bf Z}_{\bf B}$, respectively, with ${\bf D}_{\bf A}$ and ${\bf D}_{\bf B}$ diagonal.  In terms of applications, it would be interesting to develop quantitative strategies for designing and assessing policies guided by the output of the proposed model, especially with reference to the magnitude and structure of potential spillover effects. Expanding the analysis to other crime datasets from different cities would also be important to obtain a more comprehensive perspective on criminal activities in urban environments beyond Chicago.

\section{\bf Acknowledgments}
The authors are grateful to the Editor, the Associate Editor and the referees for the constructive suggestions, which helped in improving the initial version of this article. 
\\
\\
{\em Conflicts of interest: } The authors declare that they have no conflicts of interest.

\section{\bf Funding}
Carlos Díaz gratefully acknowledges financial support from the National Agency of Research and Development’s (ANID) Millennium Science Initiative Program (NCS2024\_058). Daniele Durante is funded by the European Union (ERC, NEMESIS, project number: 101116718). Views and opinions expressed are however those of the author(s) only and do not necessarily reflect those of
the European Union or the European Research Council Executive Agency. Neither the European Union
nor the granting authority can be held responsible for them.

\section{\bf Data availability}
Codes are available at \url{https://github.com/gregorzens/lr-matrix-var}.

\clearpage
\vspace{10pt}
\fontsize{9.9}{17}\selectfont

\textcolor{white}{1}

\setcounter{section}{0}
\renewcommand{\thesection}{\Alph{section}}
\numberwithin{table}{section}
\numberwithin{figure}{section}

\vspace{-8pt}
{\noindent \bf \Large Supplementary Material: ``Low-Rank Bilinear Autoregressive Models for Three-Way Criminal Activity~Tensors''}

\fontsize{9.9}{13.5}\selectfont

\vspace{8pt}
\section{\large A. Reclassification of crimes in the category  \texttt{Miscellaneous Other}} \label{sec_s1}

Table~\ref{tab:redistribute-misc-other} clarifies how crimes within the category \texttt{Miscellaneous Other} have been redistributed based on the corresponding Illinois Uniform Crime Reporting (\textsc{iucr}) code.

\begin{table}[htbp]
\centering
\footnotesize
\caption{Redistribution of  \texttt{Miscellaneous Other} (FBI Code 26) by IUCR}
\label{tab:redistribute-misc-other}
\begin{tabularx}{\textwidth}{@{}p{1cm} p{3.8cm} X p{3.8cm}@{}}
\midrule
\textbf{FBI Code} & \textbf{Target category} & \textbf{IUCR codes moved from 26} & \textbf{Notes} \\
\midrule
\texttt{09} & Arson & \texttt{1030, 1035, 5003} & Existing FBI classification \\
\texttt{16} & Prostitution & \texttt{1050} & Existing FBI classification \\
\texttt{-} & Criminal trespassing & \texttt{1330, 1335, 1350, 1360, 1365} & New  category \\
\texttt{-} & Kidnapping \& Child Harm & \texttt{1537, 1710, 1715, 1725, 1755, 1780, 1790, 1792, 4210, 4220, 4230, 4240, 4255} & New  category \\
\texttt{18} & Drug abuse & \texttt{2091, 2092, 2093, 2111, 2120, 2160} & Existing FBI classification \\
\texttt{-} & Threat \& Harassment & \texttt{2820, 2825, 2826, 3960, 3966, 3970, 3975, 3980, 4386, 4387} & New  category \\
\texttt{24} & Disorderly conduct & \texttt{2850, 2851, 2890, 2895, 3000, 3200, 3300, 3400, 3770, 3800, 3910, 3920} & Existing FBI classification \\
\texttt{05} & Burglary & \texttt{4310, 4860, 5007, 5008} & Existing FBI classification \\
\midrule
\end{tabularx}
\vspace{0.5em}
\footnotesize\emph{Notes:} Recodes apply only where the original FBI Code equals \texttt{26} (9.8\% of total criminal events).
\end{table}

\section{\large B. Additional figures} \label{sec_s2}
\vspace{-5pt}
\begin{figure}[!h]
    \centering
    \includegraphics[width=1\textwidth]{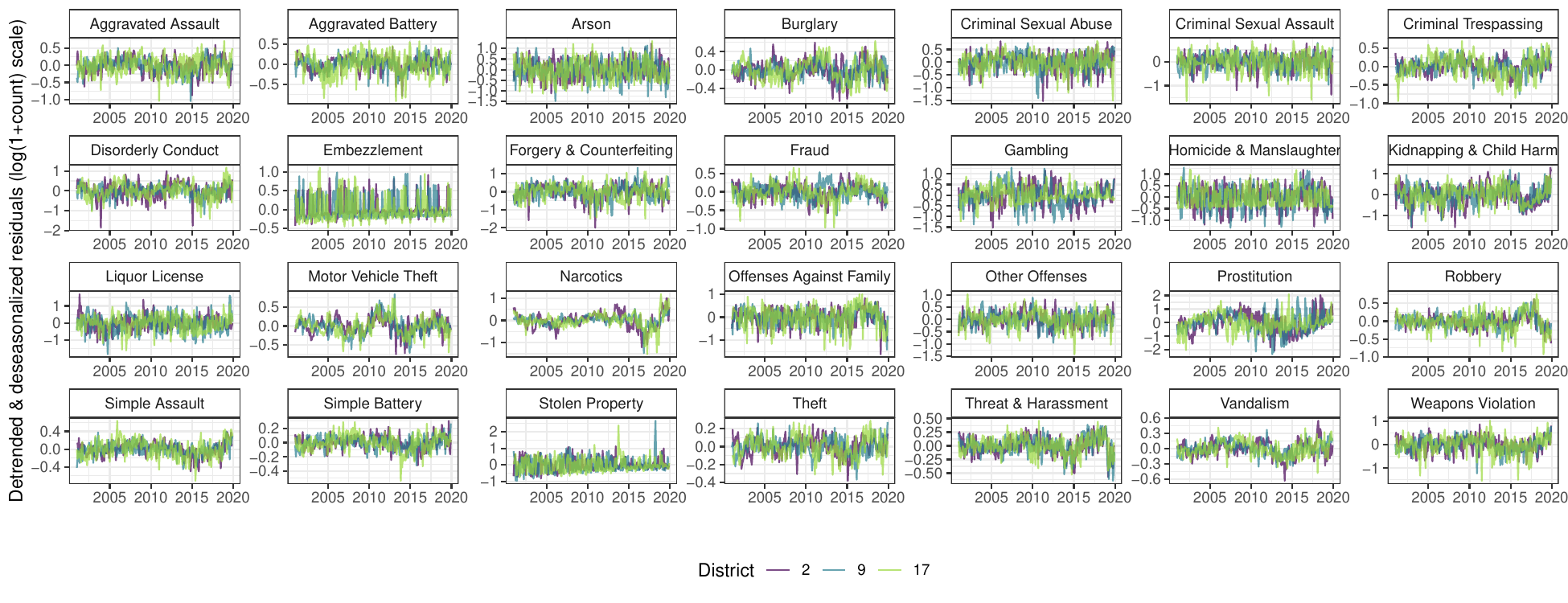}
    \vspace{-20pt}
    \caption{{For each crime category and three selected districts, the panels show the monthly detrended and seasonally adjusted series obtained by regressing the log-transformed counts on a quadratic time trend and month-of-year fixed effects (see Section 3 in the main article for details).}}
    \label{fig:log_counts_resid}
\end{figure}


\end{document}